\documentclass[reprint, superscriptaddress, 	amsmath, amssymb, aps, prb, floatfix]{revtex4-1}

\usepackage{graphicx}
\usepackage{dcolumn}
\usepackage{bm}
\usepackage{xcolor}
\usepackage{afterpage}
\usepackage{float}

\begin{document}

\preprint{APS/123-QED}

\title{Impact of Sb degrees of freedom on the charge density wave phase diagram of the kagome metal CsV$_3$Sb$_5$}

\author{Ethan T. Ritz}
\affiliation{Department of Chemical Engineering and Materials Science, University of Minnesota, Minneapolis, MN 55455, USA}
\author{Rafael M. Fernandes}
\affiliation{School of Physics and Astronomy, University of Minnesota, Minneapolis, MN 55455, USA}
\author{Turan Birol}
\affiliation{Department of Chemical Engineering and Materials Science, University of Minnesota, Minneapolis, MN 55455, USA}

\date{\today}

\begin{abstract}
 Elucidating the microscopic mechanisms responsible for the charge density wave (CDW) instability of the AV$_3$Sb$_5$ (A=Cs, K, Rb) family of kagome metals is critical for understanding their unique properties, including superconductivity. In these compounds, distinct CDW phases with wave-vectors at the $M$ and $L$ points are energetically favorable, opening the possibility of tuning the type of CDW order by appropriate external parameters. Here, we shed light on the CDW landscape of CsV$_3$Sb$_5$ via a combination of first-principles calculations and phenomenology, which consists of extracting the coefficients of the CDW Landau free-energy expansion from density functional theory. We find that while the main structural distortions of the kagome lattice in the staggered tri-hexagonal CDW phase are along the nearest-neighbor V--V bonds, distortions associated with the Sb ions play a defining role in the energy gain in this and all other CDW states. Moreover, the coupling between ionic displacements from different unit cells is small, thus explaining the existence of multiple CDW instabilities with different modulations along the c-axis. We also investigate how pressure and temperature impact the CDW phase of CsV$_3$Sb$_5$. Increasing pressure does not change the staggered tri-hexagonal CDW ground state, even though the $M$-point CDW instability disappears before the $L$-point one, a behavior that we attribute to the large nonlinear coupling between the order parameters. Upon changing the temperature, we find a narrow regime in which another transition can take place, toward a tri-hexagonal Star-of-David CDW phase. We discuss the implications of our results by comparing them with experiments on this compound.  
 
\end{abstract}

\maketitle

\section{Introduction}

The family of metallic kagome compounds AV$_3$Sb$_5$, with A~=~K, Rb, Cs, exhibit pressure-tunable superconductivity with T$_c$ ranging from 2-8 K \cite{ortiz2020cs,ortiz2021superconductivity,li2021observation, chen2021double}, coexisting with a charge density wave (CDW) order which sets in at 80-100 K \cite{ortiz2019new,ortiz2021fermi,Si2022,Zhu2022,Du2022,Consiglio2022}. Despite concerted experimental and theoretical effort \cite{review1,review2}, a complete understanding of how these electronic states change as the lattice and electronic structures are modified by pressure \cite{Chen2021Pressurea,Chen2021Pressureb,du2021pressure}, uniaxial strain \cite{Qian2021}, and doping \cite{Oey2022} remains elusive. 

Experimentally, X-ray diffraction studies at zero applied pressure show that the charge-ordered state induces a unit cell doubling along the in-plane \textit{a} and \textit{b} axes, and either a doubling or a quadrupling along the \textit{c} axis \cite{ortiz2021fermi,li2021observation,Stahl2022}. However, the exact nature of the symmetries broken by the CDW in AV$_3$Sb$_5$ as a function of temperature and pressure is still under investigation \cite{Wu2022}. For instance, several experimental results have indicated that, upon application of pressure, there is a transition between different CDW ground states, which is indirectly manifested in the double-peak structure of the superconducting dome \cite{Chen2021Pressurea,Gupta2022a,Li2022discovery}. Different CDW states have also been reported as a function of temperature and doping \cite{Wu2022,Stahl2022,li2022coexistence,kang2022charge}. Moreover, experiments have reported signatures of threefold rotational symmetry breaking either inside the CDW phase or at its onset \cite{Xiang2021twofold,zhao2021cascade,Li2022rotation,Nie2022,xu2022three}. Finally, time-reversal symmetry-breaking has been reported to coincide with the CDW transition temperature by $\mu$SR experiments \cite{mielke2022time,khasanov2022time,christensen2022loopy}, whereas Kerr rotation measurements have given conflicting results \cite{xu2022three,saykin2022_Kerr,hu2022_Kerr}.  

Theoretically, the proximity of $M$-point van Hove singularities to the Fermi level \cite{kang2022twofold} have led to the proposal that the CDW is a correlation-driven instability \cite{Denner2021,park2021electronic,Nandkishore2021,Christensen2021,review2,Kontani2022}. On the other hand, the softening of several phonon modes along the $M$-$L$ line, as seen by DFT calculations \cite{Ratcliff2021,Christensen2021,Binghai2021,Subedi2022,tsirlin2022role}, indicate the importance of electron-phonon coupling in promoting the CDW phase. Time-reversal symmetry-breaking has been interpreted in terms of loop-current patterns arising from a so-called \textit{imaginary} CDW (iCDW) instability \cite{hasan2021,park2021electronic,Nandkishore2021,feng2021chiral,Christensen2022}. Importantly, even a pure iCDW instability is expected to generally induce a ``real" CDW, making the investigation of the latter important to also shed light on the possible loop current patterns that can be established in these systems. 

Therefore, elucidating the role played by structural and electronic degrees of freedom, particularly those not associated with the V ions at the kagome layer, is important and the subject of ongoing research \cite{tsirlin2022role,jeong2022crucial,han2022orbital}. More specifically, predicting the evolution of the CDW with pressure and other tuning parameters is of critical importance to shed light on the microscopic mechanism responsible for the CDW phase, and thus to gain insight into the properties exhibited by this materials class.

In this paper, using first-principles density functional theory (DFT) in conjunction with a phenomenological Landau free-energy model, we investigate the role of the Sb degrees of freedom on the CDW instability of CsV$_3$Sb$_5$, as well as the evolution of the latter as a function of both pressure and temperature. DFT has been shown to correctly predict the CDW instabilities in this system \cite{Ratcliff2021,Binghai2021,Subedi2022,tsirlin2022role}, whose electronic correlations do not seem to be strong enough to change the lattice energetics significantly. At the same time, phenomenological Landau models for the CDW instabilities at the $M$ and $L$ points have been analyzed to reveal the possible CDW phase diagrams in these compounds \cite{park2021electronic,Christensen2021}. 

Here, we combine these two approaches by extracting the free-energy coefficients from first-principles calculations, which allows us to ``freeze'' some degrees of freedom and focus on the contributions arising from different ions. We not only find that a fourth-order Landau free energy expansion is able to capture the energy landscape of CsV$_3$Sb$_5$ with respect to its charge ordering behavior towards a staggered tri-hexagonal CDW phase, but also that the distortions of the V-bonds alone are not enough to account for the energy gain in the CDW phase. In particular, our calculations show that the apical Sb displacements are significantly more important than previously assumed for the stabilization of any of the stable CDW phases, which may have important consequences for the mechanism of the CDW. Interestingly, the relative distortions between ions of different unit cells give only a small contribution to the CDW energetics, which addresses why there are multiple CDW instabilities along the $M$-$L$ line in momentum space.  

We take advantage of the Landau free energy expansion to establish how the CDW phase evolves with pressure and temperature. We show that there is a rather narrow parameter regime in which two well separated CDW transitions (corresponding in our calculations to the staggered tri-hexagonal and tri-hexagonal Star-of-David CDW phases) onset as a function of temperature, as suggested by some experiments \cite{Wu2022,Stahl2022,khasanov2022time,guguchia2022tunable}. This result indicates that another ordered state not captured by DFT with the generalized gradient approximation, such as time-reversal symmetry-breaking iCDW order, may be necessary to account for two different transition temperatures in CsV$_3$Sb$_5$. By performing our calculations in the presence of hydrostatic pressure, we find that the staggered tri-hexagonal CDW ground state does not change, despite the fact that the $M$-point instability is absent at high enough pressures. Combined with the pressure dependence of the Landau coefficients, this suggests that the nonlinear coupling between the two CDW order parameters with wave-vectors at the $M$ and $L$ points is essential to drive the CDW instability in the kagome metals.

This paper is organized as follows: In Section~\ref{sec:methods}, we give an overview of our methods. In Subsection~\ref{ssec:zeropress}, we present the Landau free energy, and its coefficients as predicted from DFT. In Subsection~\ref{ssec:sb}, we elucidate the effect of the apical Sb ion on the phase stability of CDW. We discuss the effect of pressure in Subsection~\ref{ssec:phase_pressure}, and draw finite-temperature phase diagrams in Subsection~\ref{ssec:phase_temp}. We conclude with a summary and discussions in Section~\ref{sec:conclusions}.

\section{Methods}
\label{sec:methods}
All DFT calculations were performed using Projector Augmented Waves (PAW) as implemented in the Vienna Ab initio simulation package (VASP) version 5.4.4 \cite{kresse1993ab,kresse1996efficiency,kresse1996efficient}. We used the PBEsol exchange correlation functional, with valence configurations of 5\emph{s}$^2$5\emph{p}$^6$6\emph{s}$^1$, 3\emph{s}$^2$3\emph{p}$^6$3\emph{d}$^4$4\emph{s}$^1$, and 5\emph{s}$^2$5\emph{p}$^3$ corresponding to Cs, V, and Sb, respectively. Lattice parameters were found to be converged to within 0.001 {\AA} using a plane wave cutoff energy of 450 eV, a $\Gamma$-centered Monkhorst-Pack k-point mesh of 20$\times$20$\times$10, and a 2$^{nd}$ order Methfessel-Paxton smearing parameter of 10 meV.

Using DFT and the Landau free-energy derived in Ref. \cite{Christensen2021}, we explore the CDW behavior of CsV$_3$Sb$_5$ at various pressures. The AV$_3$Sb$_5$ compounds are predicted to have phonon instabilities at the $M$ and $L$ points of  their hexagonal Brillouin zones \cite{cho2021emergence,Qian2021}. We use the eigenvectors of the force constant matrix (FCM) corresponding to the unstable modes that transform as irreducible representations (irreps) M$_1^+$ and L$_2^-$ as a basis, and use their amplitudes as the order parameter components $M_i$ and $L_i$, respectively. At each pressure, the free-energy coefficients are found by freezing in selected combinations of the order parameters at various amplitudes and performing a least-squares fit to the data set. We confirmed that the signs and magnitudes of these coefficients do not qualitatively change when coefficients beyond fourth-order are included in the fit.

\section{Results}

\subsection{Electronic structure and free-energy coefficients at zero applied pressure and temperature} 
\label{ssec:zeropress}

At room temperature and ambient pressure, CsV$_3$Sb$_5$ adopts the $P6/mmm$ (\#191) space group, with Cs occupying the $1a$ Wyckoff site, V the $3g$ site, and Sb the $4h$ and $1b$ sites. Using DFT structural relaxation, we predict lattice parameters $a=5.424$ {\AA} and $c=9.368$ {\AA}, as well as a $z$ coordinate for the apical Sb at the $4h(1/3,2/3,z)$ site with $z=0.740$ in fractional coordinates. We find phonon instabilities that lead to CDW driven structural distortions transforming as the $M_1^+$ and $L_2^-$ irreducible representations of the space group, in agreement with earlier reports \cite{Ratcliff2021,Christensen2021}.

\begin{figure}
 \includegraphics[width=8.9cm]{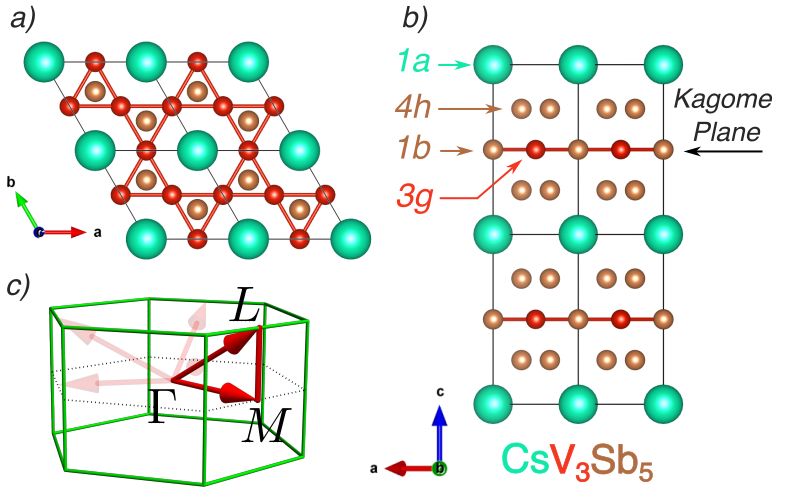}
 \caption{a) 2$\times$2$\times$2 unit cells of the high-temperature $P6/mmm$ phase of CsV$_3$Sb$_5$ along the $c$ axis. b) The same cell when viewed along the in-plane $b$ axis with the Wyckoff letters labeling each ionic site. c) The $P6/mmm$ Brillouin zone with high-symmetry points $\Gamma$, $M$, and $L$ labeled. Translucent arrows indicate the position of other vectors in the stars of $M$ and $L$.}
 \label{fig:unitcellbz}
\end{figure}

There are three distinct $M$ points and three distinct $L$ points in the hexagonal Brillouin zone, as shown in Fig. \ref{fig:unitcellbz}, which we denote as $\mathbf{k}_{Mi}$ and $\mathbf{k}_{Li}$. The $M$ point wavevectors have zero $z$ (out-of-plane) component, and they correspond to the face centers of the hexagonal Brillouin zone. $\mathbf{k}_{Li}$ have the same in-plane components as $\mathbf{k}_{Mi}$, but they also have a $z$ component of $\pi/c$, which places them at the edge centers of the hexagon on the top or bottom faces, as illustrated in Fig. \ref{fig:unitcellbz}. The different components of the $M_1^-$ and $L_2^-$ CDW order parameters correspond to displacement patterns with different wavevectors, such that the $i^{th}$ component $M_i$ and $L_i$ have the wavevectors $\mathbf{k}_{Mi}$ and $\mathbf{k}_{Li}$ respectively.

\begin{figure}
 \includegraphics[width=1.0\linewidth]{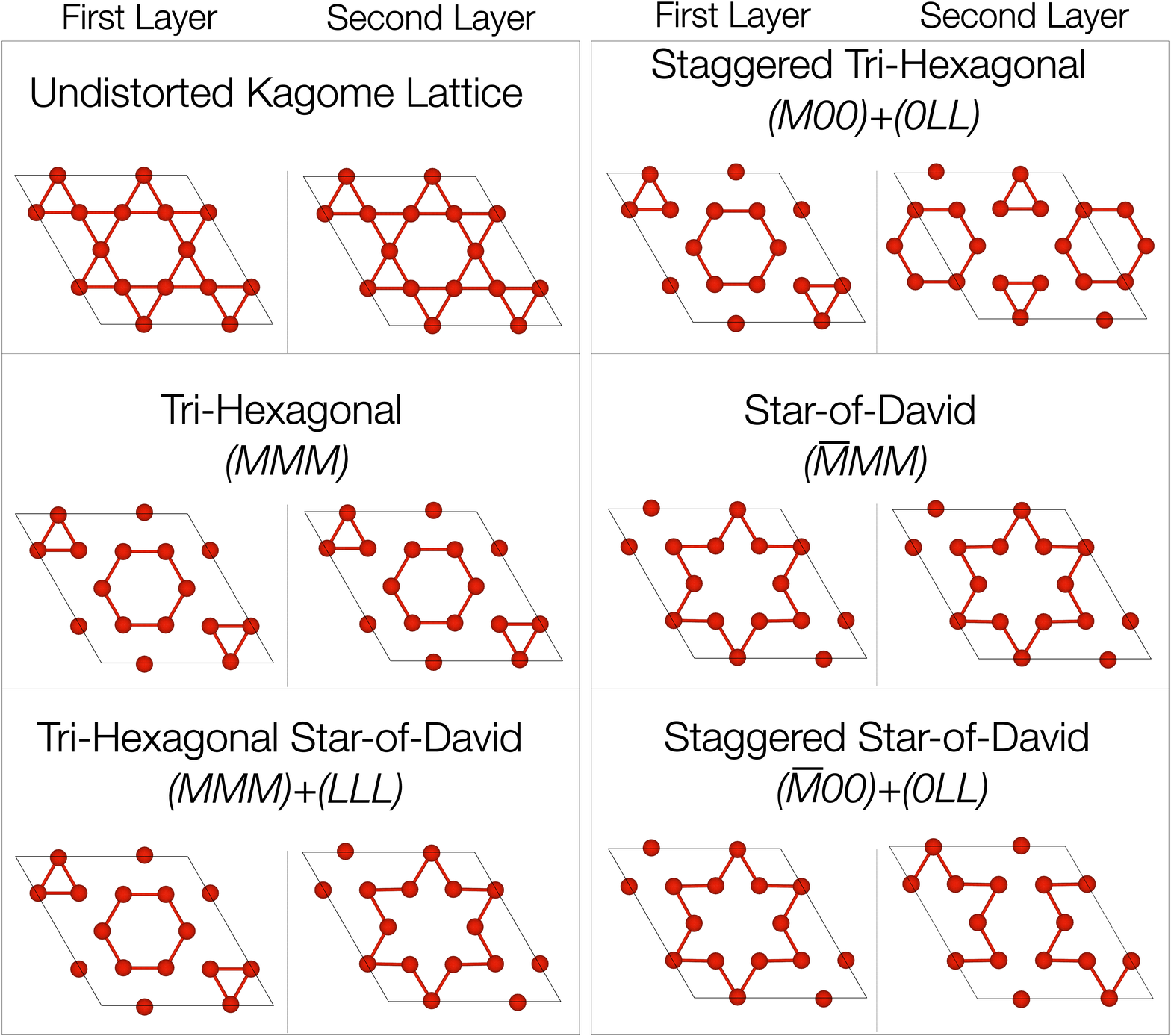}
 \caption{Illustrations of the undistorted kagome lattice and a selection of structural phases that can be reached through different directions in parameter space. The V networks are shown for two subsequent layers stacked along the $z$ axis (out of the page). The bonds shown in these panels denote relative displacement patterns. }
 \label{fig:AtomicCell}
\end{figure}

While for an isolated kagome plane there are only two types of in-plane triple-$\mathbf{Q}$ charge-order patterns (called tri-hexagonal and Star-of-David), there are several different possibilities for stacking them between consecutive layers. They correspond to distinct superpositions of the three $M_i$ and three $L_i$ CDW components, giving rise to a large number of different CDW phases with distinct symmetries \cite{Christensen2021}. In the remainder of this paper, we denote the phases reached by different directions in parameter space using the notation $(M_1M_2M_3)+(L_1L_2L_3)$, similar to the one employed in Ref. \citenum{Christensen2021}. As explained above, $M_i$ and $L_i$ refer to the amplitude of each $M_1^+$ and $L_2^-$ order parameter, respectively. For distortions where all of the $M$ or $L$ order parameters have zero amplitude, we use the notation $(L_1L_2L_3)$ and $(M_1M_2M_3)$, respectively. Illustrations of the phases discussed in this study are shown in Fig. \ref{fig:AtomicCell}. In the notation used in this figure, we take all $M$ and $L$ values to be positive, with negative values denoted by an overbar. Note that for each phase, multiple equal-energy domains that are equivalent to each other up to a translation or rotation can be obtained from combinations of order parameter components that preserve the signs of the products $M_1 \times M_2 \times M_3$ and  $L_1 \times L_2 \times L_3$. For example, $E(\overline{M}\overline{M}M)\equiv E(MMM)$. We do not discuss different domains of each phase further since they are of no consequence in a single-domain system.

Following Ref. \citenum{Christensen2021} (see also Ref. \citenum{park2021electronic}), the Landau free energy in terms of $M_i$ and $L_i$ takes the form of 
\begin{equation}
	\mathcal{F}_{tot}=\mathcal{F}_{M}+\mathcal{F}_{L}+\mathcal{F}_{ML}, 
	\label{eq:landau_tot}
\end{equation}
where
\begin{equation}
\begin{split}
\mathcal{F}_{M} =\frac{\alpha_{M}}{2}M^2+\frac{\gamma_M}{3}M_1M_2M_3+\frac{u_M}{4}M^4\\+\frac{\lambda_M}{4}\left(M_1^2M_2^2+M_1^2M_3^2+M_2^2M_3^2 \right),
    \end{split}
     \label{eq:landau_M}
\end{equation}
\begin{equation}
\begin{split}
\mathcal{F}_{L}=\frac{\alpha_{L}}{2}L^2 +\frac{u_L}{4}L^4+\frac{\lambda_L}{4}\left(L_1^2L_2^2+L_1^2L_3^2+L_2^2L_3^2 \right),
    \end{split}
    \label{eq:landau_L}
\end{equation}
and
\begin{equation}
\begin{split}
\mathcal{F}_{ML}=\frac{\gamma_{ML}}{3}\left( M_1L_2L_3+L_1M_2L_3+L_1L_2M_3 \right) \\+ \frac{\lambda^{(1)}_{ML}}{4} \left(M_1M_2L_1L_2+M_1M_3L_1L_3+M_2M_3L_2L_3 \right)
\\ +   \frac{\lambda^{(2)}_{ML}}{4}\left(M_1^2L_1^2+M_2^2L_2^2+M_3^2L_3^2 \right) +\frac{\lambda_{ML}^{(3)}}{4}M^2L^2. 
\end{split}
\label{eq:landau_mixed}
\end{equation}
Here, we defined $M^2=M_1^2+M_2^2+M_3^2$ and $M^4=(M^2)^2$, with $L^2$ and $L^4$ defined analogously. The form of this free energy is determined by symmetry, and it is the most general fourth-order polynomial of $L_i$ and $M_i$ that transforms as a scalar under the symmetry operations of space group $P6/mmm$. A necessary, but not sufficient, condition for a term to appear in this expansion is that its total wavevector is zero, so that the free energy is invariant under lattice translations. The sum of the three $M$ point wavevectors is zero modulo a reciprocal lattice vector ($\mathbf{k}_{M1}+\mathbf{k}_{M2}+\mathbf{k}_{M3}=0$), and hence the trilinear term $\gamma_M M_1M_2M_3$ is allowed
\footnote{We use the word `trilinear' to refer to expressions which involve three different order parameter components to differentiate them from the more common cubic terms such as $M^3$. This is a more general use the term than, for example, in the ferroelectrics community where the word `trilinear' is used to refer to a combination of three different irreps (${\propto}P_1 Q_1 R_3$, etc.). For example, see Ref.~[\onlinecite{Etxebarria2010, Mulder2013, Li2020Barrier}].}. 
However, there is no trilinear term proportional to $L_1L_2L_3$ because $\mathbf{k}_{L1}+\mathbf{k}_{L2}+\mathbf{k}_{L3}\neq0$. Accordingly, there are trilinear terms between two $L_i$ and one $M_i$ component proportional to $M_1L_2L_3$ because $\mathbf{k}_{M1}+\mathbf{k}_{L2}+\mathbf{k}_{L3}=0$.

\begin{center}
\begin{table*}[]
    \centering
    \begin{tabular}{|c|r|r|r|r|r|r|r|r|r|r|r|}
        \hline
   Pressure &     $\alpha_M$ & $\alpha_L$ & $\gamma_M$ & $\gamma_{ML}$ & $u_M$ & $u_L$ & $\lambda_M$  & $\lambda_L$ & $\lambda^{(1)}_{{ML}}$ & $\lambda^{(2)}_{{ML}}$ & $\lambda^{(3)}_{{ML}}$ \\
 (GPa)& (eV/\AA$^2$) & (eV/\AA$^2$) & (eV/\AA$^3$) & (eV/\AA$^3$) & (eV/\AA$^4$) & (eV/\AA$^4$) & (eV/\AA$^4$)  & (eV/\AA$^4$) & (eV/\AA$^4$) & (eV/\AA$^4$) & (eV/\AA$^4$) \\
        \hline \hline
        0.00 & {-2.53} & -2.92 & -22.16 & -24.15 & 76.43 & 89.93 & -137.67 & -194.47 & 347.73 & 332.28 & 24.20 \\
        0.80 & {-1.78} & -2.42 & -22.88 & -30.15 & 73.05 & 77.68 & -128.01 & -127.43 & 347.38 & 341.12 & 29.53 \\
        1.74 & {-1.00} & -2.21 & -17.57 & -27.78 & 52.87 & 74.91 & -94.82 & -122.71 & 244.04 & 324.52 & 24.79 \\
        2.50 & {-0.53} & -1.31 & -10.60 & -32.66 & 70.71 & 82.46 & -118.16 & -138.51 & 266.06 & 333.90 & 12.58 \\
        3.50 & { 0.51} & -0.89 & { -7.06} & -31.62 & 16.01 & 110.61 & 61.26 & -203.07 & 98.22  & 353.52 & 3.45 \\
        5.00 & { 1.22} & -0.43 & {  3.56} & -20.41 & 101.42 & 181.22 & -64.76 & -279.73 & 400.84 & 639.46 & -32.88 \\
        \hline
    \end{tabular}
    \caption{Coefficients of the Landau free energy corresponding to Eqs. \ref{eq:landau_M}, \ref{eq:landau_L}, and \ref{eq:landau_mixed} obtained from fits to DFT data for different applied pressures. The units for all second-order ($\alpha$), third-order ($\gamma$), and fourth-order ($u$ and $\lambda$) coefficients are eV/{\AA}$^2$, eV/{\AA}$^3$, and eV/{\AA}$^4$, respectively, defined per 2$\times$2$\times$2 (eight formula unit) supercell, which is the smallest supercell commensurate with all three wavevectors in the stars of $M$ and $L$.}
    \label{tab:constants}
\end{table*}
\end{center}

The Landau coefficients (Greek letters and $u$) in Eq.~\ref{eq:landau_mixed} are material-specific coefficients that can be obtained from DFT. We calculate these quantities by performing a simultaneous least-squares fit of the total energy extracted from DFT as a function of a selected combination of order parameters \footnote{All numerical fits were performed using the SciPy Python library. In order to insure a good fit to the 11 coefficients in Equations \ref{eq:landau_M}-\ref{eq:landau_mixed}, the energies associated with distortions in 13 directions in parameter space were calculated, then simultaneously fit using the non-linear least-squares method. These directions were selected to insure that the resulting system of least-squares normal equations spanned the space of all unknown coefficients, and consist of $(L00)$, $(LL0)$, $(M00)+(0LL)$, $(M00)+(LL0)$, $(MM0)$, $(M00)$, $(MM0)+(LL0)$, $(MMM)$, $(MMM)+(L00)$, $(MMM)+(LLL)$, $(M00)+(\frac{L}{2}LL)$, and $(\frac{M}{2}00)+(LLL)$. A Landau free energy including terms up to 5$^{th}$ order showed no appreciable change in the 2$^{nd}$- through 4$^{th}$-order coefficients. Note that for the 3.50 GPa pressure data, the M-point phonons are stable ($\alpha_M>0$), but just barely so. This makes the energy surface highly sensitive to distortions of pure $M_1^+$ order, and leads to difficulties in distinguishing between even-order terms in the energy expansion that involve $M_1^+$ distortions only. Thus, the quantitative accuracy of $\alpha_M$, $u_M$, and $\lambda_M$ at this pressure value may be less robust than for other values in Table \ref{tab:constants}}. In practice, we take the eigenvectors associated with each of the unstable eigenvalues of the force constant matrix that transform as $M_1^+$ and $L_2^-$ as the distortions associated with the $M$ and $L$ point order parameters, respectively. Then, we use DFT to calculate the energy associated with combinations of these distortions with various amplitudes and frozen into a 2$\times$2$\times$2 supercell commensurate with the stars of both the $M$ and $L$ points in reciprocal space. The coefficients obtained from these calculations at different values of hydrostatic pressure are shown in Table \ref{tab:constants}.

Since $\alpha_L$ and $\alpha_M$ are proportional to the square of the frequencies of the unstable phonon modes at $L$ and $M$, respectively, these results show that at 3.50 GPa, the $M_1^+$ CDW instability has disappeared, whereas the $L_2^-$ CDW instability is suppressed at approximately 5.00-6.50 GPa. We note that the exact value of pressure where this suppression occurs is hard to pinpoint from DFT, as it also depends on the choice of exchange-correlation functional and van der Waals corrections. Nevertheless, the qualitative trends are expected to be reliable \cite{Gupta2022a}. 

The ``pure $M$'' ($\gamma_M$) and ``mixed irrep'' ($\gamma_{ML}$) trilinear coupling coefficients are significant at almost all pressures, and hence play an important role in determining the structure of the energy surface. Also of interest are the biquadratic terms $\lambda_M$ and $\lambda_L$, which take \emph{negative} values at all pressures explored. While other fourth-order terms ensure positive curvature at large displacement amplitudes that keep the free energy bounded, and the third-order terms are dominant in CsV$_3$Sb$_5$, $\lambda_L$ and $\lambda_M$ could play an important role in shaping the competition between CDW phases in other kagome systems, where their magnitude could be larger.

\begin{figure*}
\includegraphics[width=1.0\textwidth]{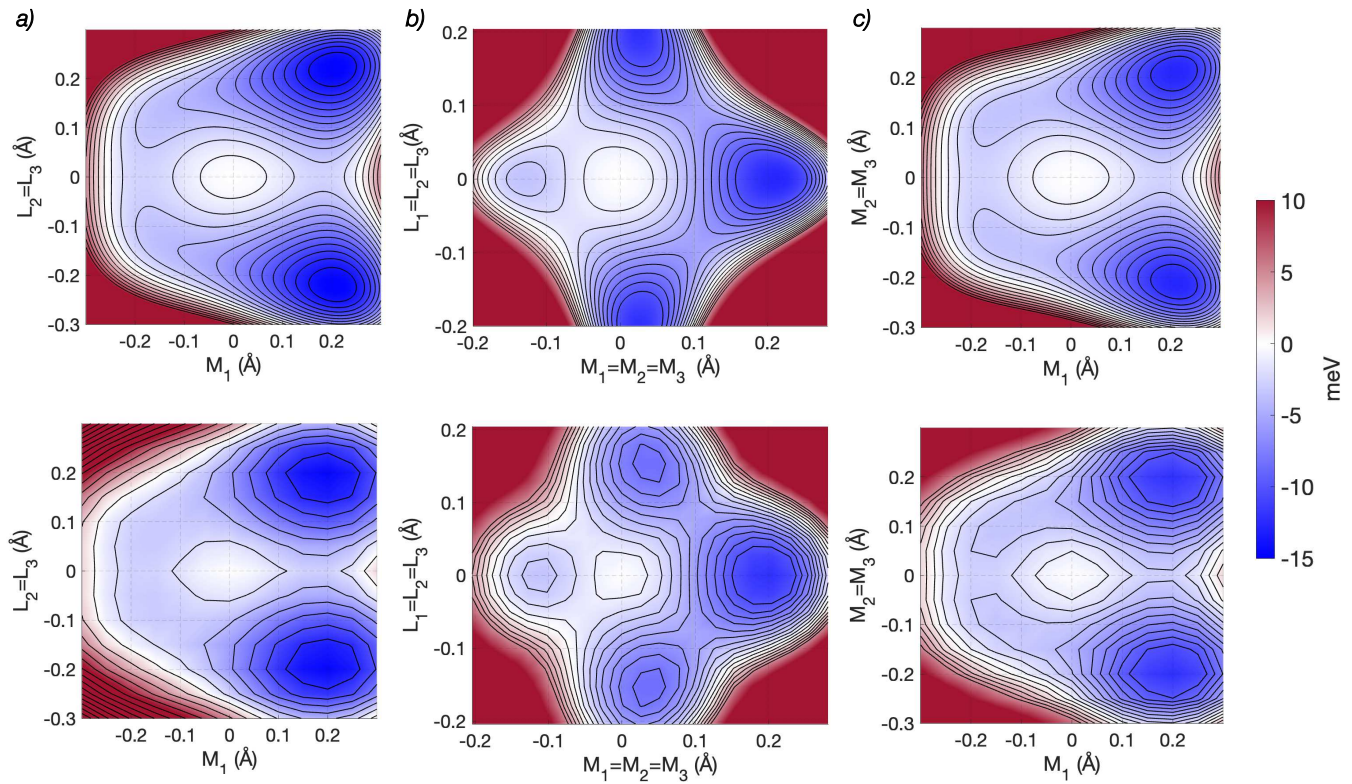}
	\caption{Energy surfaces along specific cuts in the six-dimensional phase space $(M_1, M_2, M_3, L_1, L_2, L_3)$ obtained from the Landau free-energy fit (top panels) and from DFT (bottom panels). (a),(b), and (c) correspond, respectively, to the order parameter subspaces spanned by $(M_1, M_2=M_3=0)$ and $(L_1=0, L_2 = L_3)$; $(M_1=M_2=M_3)$ and $(L_1=L_2=L_3)$; $(M_1, M_2=M_3)$ and $(L_1=L_2=L_3=0)$. }
   \label{fig:EnergyContours}
\end{figure*}

We show the energy contour plots along various cuts of the six-dimensional phase space $(M_1, M_2, M_3, L_1, L_2, L_3)$ in Fig.~\ref{fig:EnergyContours}. In order to assess whether the fourth-order expansion is sufficient to capture the topography of the energy landscape, we show the energies calculated using the Landau free-energy expression and the coefficients from Table~\ref{tab:constants} on the top panels of the figure, whereas the bottom panels display the energies as obtained directly from DFT.
The comparison reveals good qualitative and quantitative agreement between the fit and DFT, with the fitted free energy capable of capturing all local minima, as well as their relative amplitudes and energies. This suggests that a fourth-order free-energy expression is sufficient to capture the energetics of the CDW degrees of freedom. At zero temperature and zero applied pressure, a minimization of the free energy predicts the equilibrium phase to be the $(M00)+(0LL)$ (staggered tri-hexagonal) phase, in agreement with our DFT structural relaxations, which take into account strain degrees of freedom as well. Earlier first principles studies also reported the same ground state structure from structural relaxations \cite{Ratcliff2021,Binghai2021}.

These energy surfaces provide insights about various unusual features of the coupling between the CDW order parameters. 
Fig.~\ref{fig:EnergyContours}a is very asymmetric as a function of $M_1$, which is solely due to the trilinear coupling term $\gamma_{ML}$. This term breaks the symmetry between $\mp M_1$ in this figure, and favors $M_1>0$, which amounts to breaking the degeneracy between the staggered Star-of-David phase, $(\overline{M}00)+(0LL)$, and the staggered tri-hexagonal phase, $(M00)+(0LL)$, by favoring the latter. 
Another interesting feature, visible in Fig.~\ref{fig:EnergyContours}b, is that the pure $(MMM)$ phase (the minimum on the horizontal axis) is lower in energy than the mixed $(MMM)+(LLL)$ phase (the minimum slightly off of the vertical axis), despite the latter being a subgroup of the former. This is due to an interplay between third-order and fourth-order terms in the free energy expansion: The $(MMM)$ phase has a considerable energy gain from the pure trilinear coupling $\gamma_M$, whereas the $(MMM)+(LLL)$ phase only has a small gain from the mixed trilinear coupling $\gamma_{ML}$ because of the small amplitude of $M$. A larger value of $M$ in the $(MMM)+(LLL)$ phase is disfavored because of the fourth-order couplings $\lambda^{(i)}_{ML}$, which results in the relative stability of the $(MMM)$ phase over $(MMM)+(LLL)$. 

The similarity between Figs.~\ref{fig:EnergyContours}a and c is also striking. For Fig.~\ref{fig:EnergyContours}c, where only CDW orders at the $M$-point are considered, the $\gamma_M$ trilinear term drives the minimum of the free energy to a point in phase space where $|M_1|=|M_2|=|M_3|$, regardless of the sign of $\gamma_M$, which only determines whether the Star-of-David ($M<0$) or the tri-hexagonal ($M>0$) phase is favored \cite{Christensen2021}. Both of these phases retain the six-fold rotational symmetry in their point group, which would be broken if any of the $|M_i|$ were not equal to the others. However, Fig. \ref{fig:EnergyContours}a shows an energy minimum very close to $|M_1|=|L_2|=|L_3|$, despite the fact that no similar symmetry conditions or constraints are enforced by the form of the free energy. 

\begin{figure}
 \includegraphics[width=8.9cm]{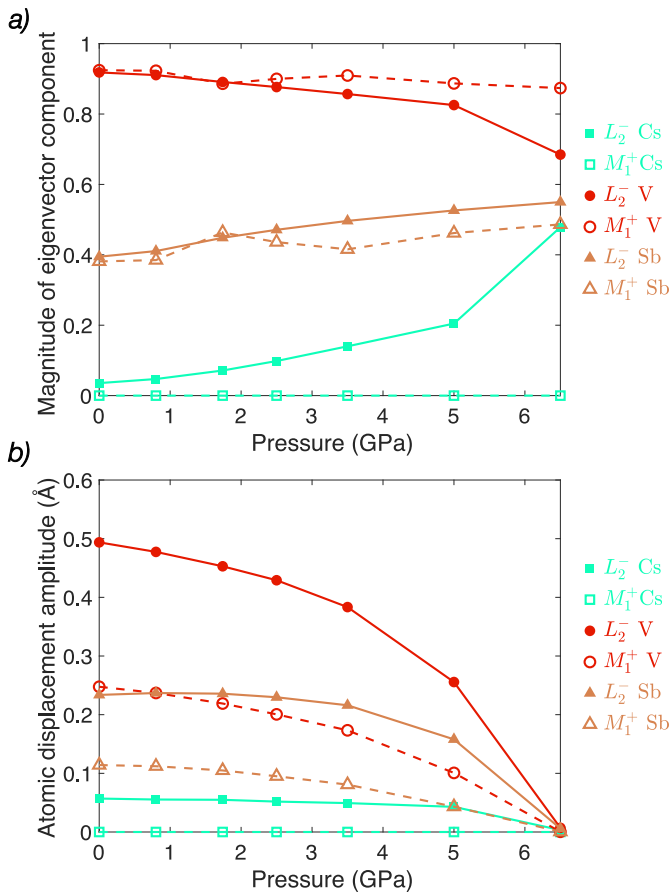}
 \caption{a) Norm squared magnitude of the displacements associated with the eigenvectors of the force constant matrix that correspond to the $M_1^+$ and $L_2^-$ CDW instabilities decomposed by ionic species. The total magnitude is normalized to 1. b) Total amplitude of the ionic displacements associated with the order parameter components $L$ and $M$ for the $(M00)+(0LL)$ phase obtained from DFT and calculated with the ISODISTORT program.}
 \label{fig:atomicdist}
\end{figure}

\begin{figure}
 \includegraphics[width=8.9cm]{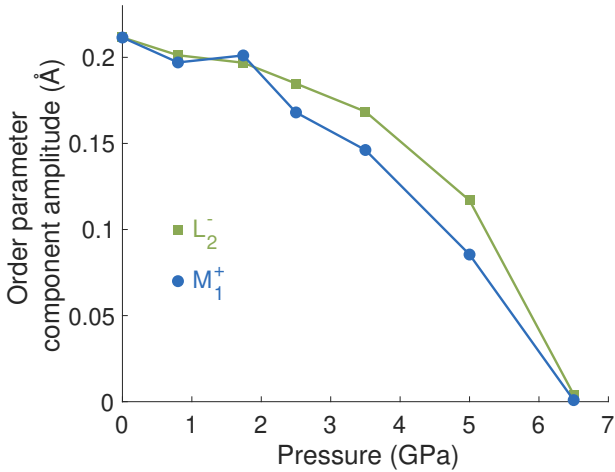}
	\caption{Total amplitude of the order parameter components $L$ and $M$ for the $(M00)+(0LL)$ phase relaxed with DFT. At each pressure, the ISODISTORT program was used to decompose the distortion of the relaxed cell with respect to the high-symmetry kagome structure into contributions from the $L_2^-$ and $M_1^+$ irreps.  Since two equivalent and nonzero $L$ order parameter components appear in the $(M00)+(0LL)$ phase, the value for $L_2^-$ plotted in the figure is equal to the total value reported by ISODISTORT scaled by $\frac{1}{\sqrt{2}}$. }
 \label{fig:scaledSAM}
\end{figure}

We explored the structural details of the tri-hexagonal CDW $(M00)+(0LL)$ phase, with space group $Fmmm$  and which is the global minimum of the free energy, using DFT to relax all structural degrees of freedom in it. We then used ISODISTORT \cite{stokes_iso,stokes2006isodisplace} to decompose the distortions from the parent kagome structure in terms of the irreps of space group $P6/mmm$. The relaxed structure hosts multiple distortions with different irreps, including $\Gamma$-point strain modes. These uniform distortions are induced by their coupling to the $L_2^-$ and $M_1^+$ CDW distortions. For simplicity, we ignore them and focus only on the unstable $L_2^-$ and $M_1^+$ modes. We find that the total distortion can indeed be described by a superposition of $L_2^-$ and $M_1^+$ irreps of near-equal magnitudes (within 1\% at zero pressure, see Fig. \ref{fig:scaledSAM}), which primarily takes the form of distortions of the V and Sb ions as seen from the decomposition of each irrep into separate ionic displacements (Fig.~\ref{fig:atomicdist}, discussed further in Section \ref{ssec:sb}). Note that neither the $L_2^-$ or $M_1^+$ irreps accommodate distortions of the Sb ions located in the kagome plane (Wyckoff site $1b$ in space group $P6/mmm$), nor are there any degrees of freedom in the Wyckoff sites corresponding to those ions in the $(M00)+(0LL)$ CDW phase (sites $4a$, $4b$, $8e$ of space group $Fmmm$). This means that all Sb distortions in Fig. \ref{fig:atomicdist} refer to the ``apical'' (Wyckoff site $4h$ in $P6/mmm$) Sb ions located above and below the kagome plane.  

Taken together, these results suggest that, while the coupling between the $M$-point and $L$-point CDWs play an important role in determining the global energy minimum, it is \emph{intralayer} interactions between ions in the same vanadium kagome sublattice, as well as their coupling to neighboring Sb ions, that dominates the behavior of the energy surface, with little coupling to layers in neighboring unit cells. This is supported by the similarity of pure-$L$ and pure-$M$ coefficients in Table \ref{tab:constants} at 0.0GPa, especially the $\alpha$ terms, which are proportional to the squared phonon frequencies and are within about 15\% of each other. Since these values are so close, we conclude that the vibrational frequencies are fairly agnostic as to whether their associated distortions are in-phase or out-of-phase with respect to neighboring kagome layers, again supporting a physical picture of weak interaction between nearest-neighbor unit cells. 

While this interpretation is consistent with the layered crystal structure of AV$_3$Sb$_5$ and is also in line with many studies that assume well isolated vanadium kagome layers that are stacked with some space-filling ions in between, in the following sections we argue that ions other than vanadium are also important for the CDW phase.

\begin{figure*}
\includegraphics[width=1.0\textwidth]{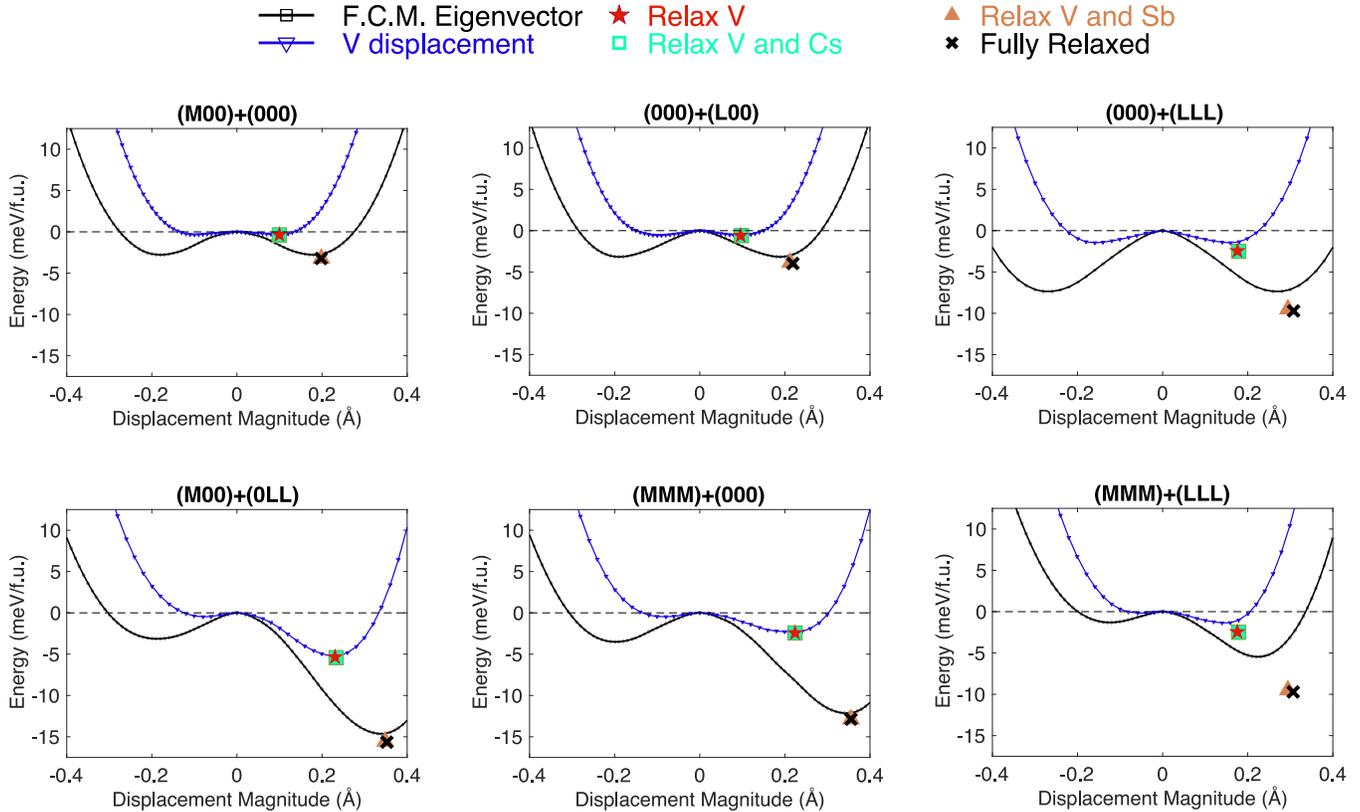}
	\caption{Comparison of the effects of different ionic degrees of freedom on the energy surface for various CDW distortion patterns. The black lines overlaid with square markers correspond to the energy change associated with a distortion constructed from the force-constant matrix (FCM) eigenvectors associated with the indicated direction in parameter space. These eigenvectors involve displacement of not only V, but also of apical ($1b$) Sb, and of Cs (for $L_2^-$ order parameters only). Blue lines overlaid with inverted triangles correspond to distortions constructed using only V-ion displacements, leaving all other degrees of freedom fixed. Using the minimum of this blue curve as initial conditions, the results of constrained structural relaxations are shown as additional symbols. Red stars correspond to letting only V ions relax further, while green squares correspond to letting V and Cs ions relax. Brown triangles correspond to letting V and Sb ions relax, and black crosses, to allowing all internal degrees of freedom to relax. Note that for all directions in the CDW parameter space spanned by the $M$ and $L$ order parameter components, the magnitudes of $M$ and $L$ were selected to be equal. While these directions are useful for illustrating the energy surfaces, the magnitudes of $L$ and $M$ need not be equal in general, and the values of $L$ and $M$ which minimize the total energy do not necessarily lie along a direction where $|M|=|L|$.}
   \label{fig:dofs6phases}
\end{figure*}

\subsection{Importance of apical Sb displacement} \label{ssec:sb}

While much of the theoretical work on AV$_3$Sb$_5$ has focused on the role of the V ions that form the kagome lattice, there is a growing body of evidence that points to the important role of the Sb ions \cite{tsirlin2022role,jeong2022crucial,han2022orbital,Oey2022}. Importantly, the $p_z$ orbitals of both types of Sb anions (in-plane and apical) together give rise to a $\Gamma$-centered Fermi surface pocket \cite{tsirlin2022role,jeong2022crucial,Oey2022}. Moreover, certain types of imaginary CDW or loop current phases are predicted to induce magnetic moments on the in-plane Sb ions \cite{Christensen2022}. As for the apical Sb ions, constrained random phase approximation calculations report that they give an important contribution to the correlation strength in CsV$_3$Sb$_5$ \cite{jeong2022crucial}. Finally, as discussed in the previous section, both $L$ and $M$ CDW order parameters involve displacements of the apical Sb ions, while the planar Sb ions remain fixed by symmetry in many CDW phases.  

In addition to their role in the electronic structure, the Sb ions can also be important in determining the relative stability of different CDW phases. Indeed, while the ionic displacements associated with the unstable $M_1^+$ and $L_2^-$ modes obtained from phonon calculations are primarily dominated by the V ions, decomposition of force constant matrix (FCM) eigenvectors and the relaxed structures suggests that the displacements of other ions may play an important role in the energetics of the CDW phase. 
In Fig. \ref{fig:atomicdist}a, we plot the components of the FCM eigenvectors of the unstable $M_1^+$ and $L_2^-$ modes. The FCM eigenvectors are conceptually similar to the dynamical matrix eigenvectors, which give the displacement pattern of phonon modes when multiplied by the square roots of the atomic masses. However, for unstable lattice modes, the FCM eigenvectors give a more precise description of the displacement pattern since they do not depend on the masses of the atoms. The FCM eigenvector of both $M_1^+$ and $L_2^-$ modes at zero pressure has the largest contribution from the V ions, with Sb atoms having a smaller total displacement despite the fact that there is a larger number of Sb ions being displaced. Under pressure, the most significant change is an increase in the out-of-plane Cs ion displacement in the $L_2^-$ mode, but the Sb contribution remains relatively flat. 

The FCM eigenvectors only contain information about the nature of the instabilities in harmonic order, and cannot capture the effects of the higher order terms in the lattice Hamiltonian. As a result, they cannot explain which ions' displacements lead to the largest energy gain, because the higher order interactions can give rise to a different ionic displacement pattern when the lattice is relaxed. In order to evaluate this possibility, in Fig. \ref{fig:atomicdist}b, we plot the total amplitudes of ionic displacements in the staggered tri-hexagonal $(M00)+(0LL)$ phase decomposed into different irreps and ions. We note that the Sb displacements with $L_2^-$ character in the relaxed structure are as large as the V displacements with $M_1^+$ character, and therefore, Sb displacements may be responsible of a significant energy gain in the CDW phase. 

To investigate the importance of the degrees of freedom other than V displacements, we perform a series of DFT calculations, the results of which are shown in Fig. \ref{fig:dofs6phases}. Each panel of Fig.~\ref{fig:dofs6phases} shows the energy calculated for a different CDW phase [$(M00)$, $(MMM)$, etc.] as a function of ionic displacements. For each phase, we consider two different displacement patterns: The black solid lines correspond to distortions according to the FCM eigenvectors, and thus include a displacement of Sb and Cs ions in addition to the V ions. The blue solid lines, on the other hand, are the energies when only the V ions are displaced according to the same pattern, but the positions of the Cs and Sb ions are kept fixed. 
In every case, we find that that black curve goes much deeper than the blue one. In other words, even though there is an energy gain even when only V ions are displaced, 
this energy gain is much less than that obtained when all ions are displaced. 
Thus, the \emph{greatest share} of the free energy change is associated with degrees of freedom \emph{other} than the vanadium displacement. These other degrees of freedom are not unstable by themselves, however, it is their interactions with the V displacements (i.e. the off-diagonal elements of the FCM) that lead to these large energy gains.

In order to further learn how these other degrees of freedom contribute to free energy, we also performed constrained ionic relaxations in DFT. In these calculations, the ionic positions are allowed to relax to the minimum energy configuration that preserves symmetry. As the starting point, we used the minimum of each V--V-only distortion pattern (i.e. the minimum of the blue curves), and performed four different types of relaxations while keeping certain ions' positions fixed. The results of these calculations are shown by different symbols in Fig.~\ref{fig:dofs6phases}. 
Keeping the Cs and Sb ions fixed while allowing the V ions to relax (red stars) gives rise to a marginal energy gain and additional displacements compared to the minimum of the blue curve. This indicates that, in this relaxed structure, the V displacements differ very little from the pattern obtained from the force constant matrix eigenvectors. 
Relaxing both the V and the Cs ions (green squares) makes almost no difference either, implying that the Cs ions have almost no effect in the stabilization of the CDW phase. Hence, we conclude that any difference between the CDW behavior in CsV$_3$Sb$_5$, RbV$_3$Sb$_5$ and KV$_3$Sb$_5$ is likely due to the size effects of the different alkali metals, which change the lattice parameter and hence the electronic structure, and not directly due to steric effects related with the alkali metal displacements.

Unlike the Cs ions, the relaxation of the Sb ions make a significant difference in the amount of energy gained. Relaxing both the V and the Sb ions (brown triangles) leads to both an energy and a mode amplitude that are close to the minimum of the curve where all ions are displaced according to the FCM eigenvector. Therefore, while the V-ion degrees of freedoms by themselves can explain the presence of the CDW instability and qualitatively capture the ground state symmetry, the displacements of the Sb ions are essential in obtaining the large energy gain that stabilizes the CDW phase at ${\sim}80$~K. 
Finally, also relaxing Cs together with V and Sb (black crosses) makes very little difference, confirming the unimportance of Cs displacements.

\subsection{Effect of Hydrostatic Pressure} \label{ssec:phase_pressure}

CsV$_3$Sb$_5$ exhibits a rich phase diagram under hydrostatic pressure, where the CDW order is suppressed around 2~GPa, while superconductivy displays a double-peak dome behavior when it coexists with CDW \cite{Yu2021b, Zhang2021Pressure, Chen2021Pressurea, Chen2021Pressureb, Du2022, Gupta2022a, Qian2021}. In this subsection, we study the evolution of the CDW under pressure by focusing on the coefficients of the Landau free-energy to predict any change in the CDW ground state of the system. 

In Table~\ref{tab:constants}, we show the values of the free energy coefficients for CsV$_3$Sb$_5$ under pressure.  As the pressure increases, both quadratic coefficients $\alpha_M$ and $\alpha_L$ become less negative, which is consistent with the disappearance of the CDW under pressure in the experiments discussed above. Interestingly, $\alpha_M$ changes sign and becomes positive while $\alpha_L$ remains negative, indicating that, in the higher pressure regime, it is the $L$ instability that drives the CDW transition. However, as shown in Figure \ref{fig:scaledSAM}, the $M_1^+$ distortions continue to contribute significantly to the CDW phase, even in this regime. 

\begin{figure} \includegraphics[width=0.7\linewidth]{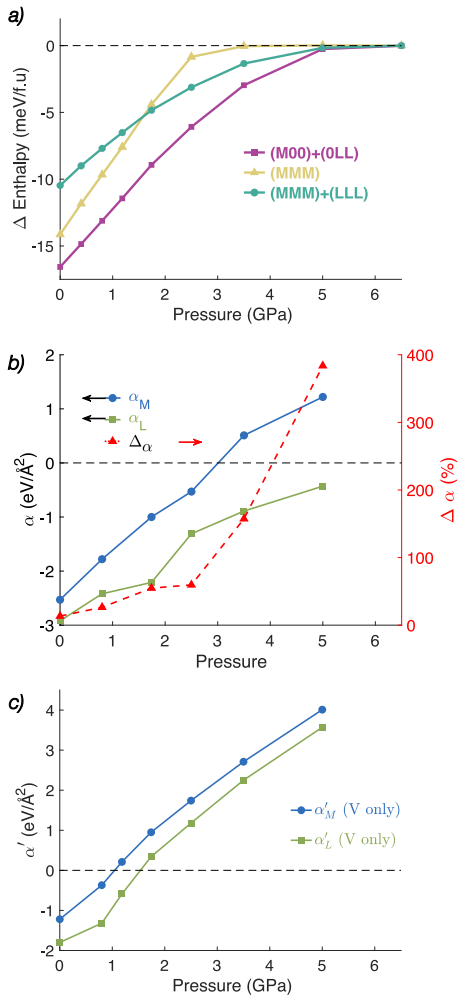}
	\caption{(a) The difference in enthalpy per formula unit between each of the indicated CDW phases and the undistored kagome lattice as a function of pressure. Only the three lowest enthalpy phases predicted by DFT are shown.
	(b) Quadratic (left axis) coefficients of the Landau free energy and $\Delta_{\alpha}=\frac{\alpha_L-\alpha_M}{\alpha_L}$ (right axis) as a function of pressure. 
	(c) Pressure-dependence of the coefficients $\alpha^\prime_M$ and $\alpha^\prime_L$. They correspond to the $\alpha_M$ and $\alpha_L$ of Eqs. \ref{eq:landau_M} and \ref{eq:landau_L} calculated for a distortion pattern transforming as $M_1^+$ and $L_2^-$, respectively, but involving only the displacements of the V ions along nearest-neighbor V--V bonds. }
 \label{fig:pressure}
\end{figure}

We also find large changes in the the third-order coefficients $\gamma_M$ and $\gamma_{ML}$, with $\gamma_M$ becoming less negative much faster than $\gamma_{ML}$ with increasing pressure, and even changing sign near 5.0 GPa. Despite these changes, the $(M00)+(0LL)$ CDW phase is predicted by both DFT and the fitted free energy function to be the lowest enthalpy structure throughout the entire pressure range, as seen in Fig.~\ref{fig:pressure}a. However, the energy difference between the three lowest energy CDW phases shown in Fig.~\ref{fig:pressure}a is typically of the order of few meV per formula unit, which suggests that the vibrational entropy of the ions, which is not taken into account in our calculations, can be large enough to lead to a different ground state or even a phase transition under pressure. 
We also note that the critical pressure we predict for the complete suppression of the CDW, $P = 6.5$~GPa, which we reported previously in Ref. \cite{Gupta2022a}, is higher than the experimentally reported value, which may in large part be due to the systematic errors in DFT and GGA approximations. The trends in the coefficients, which do not sensitively depend on the equilibrium lattice constants, are nevertheless reliable.

As discussed in Section \ref{ssec:zeropress}, the similarity between $\alpha_M$ and $\alpha_L$ can be interpreted as a measure of the weakness of the coupling between the displacements of ions in neighboring unit cells along the $c$ axis. Moreover, the CDW structural distortion patterns associated with the unstable $L_2^-$ modes are nearly the same as those associated with the $M_1^+$ modes, except that in the $L_2^-$ mode each kagome layer is out of phase from its neighbors \cite{Christensen2021}. Thus, if the frequencies of the $L_2^-$ and $M_1^+$ phonon modes are close in value, it suggests that the frequencies are relatively indifferent to the relative phases between each layer. Since $M$ and $L$ are the two endpoints of the $U$ line in reciprocal space parameterized by $(\frac{\pi}{a}, \frac{\pi}{a}, q_z)$, the similarity of $\alpha_M$ and $\alpha_L$ can be used as a rough measure of the sensitivity of a phonon frequency to $q_z$, which modulates the phase relation between neighboring layers. Fig.~\ref{fig:pressure}b shows how the relative difference between the quadratic coefficients of the $M$ and $L$ modes, $\Delta_{\alpha}=\frac{\alpha_L-\alpha_M}{\alpha_L}$, changes as a function of pressure. $\alpha_M$ becomes positive faster than $\alpha_L$, and as the latter approaches zero, $\Delta_\alpha$ increases sharply. This steady increase with pressure indicates an increased importance of the interlayer coupling under pressure. This may be due to the decrease of the out-of-plane $c$ axis, enhancing the possibility that the leading instability transforms like an irrep associated with another wave-vector on the $U$ line.

We also find that the the non-vanadium degrees of freedom discussed in Section \ref{ssec:sb} play a critical role in the evolution of the structural instability with pressure. Fig.~\ref{fig:pressure}c shows the $\alpha'_M$ and $\alpha'_L$ values calculated using the distortion patterns associated only with V displacements rather than the eigenvectors of the force constant matrix. These are not only far less negative than the actual $\alpha$ values show in Figure \ref{fig:pressure}b, but also become positive at a much smaller pressure. As a result, in the absence of Sb and Cs displacements, the CDW instabilities are greatly weakened, in line with the results in Fig.~\ref{fig:dofs6phases}. Thus, the displacements of Sb ions not only play an important role in the stability of the CDW phase of CsV$_3$Sb$_5$ at zero pressure, but also in controlling how the CDW instability evolves with pressure.

\subsection{Finite Temperature Phase Diagrams}
\label{ssec:phase_temp}

While the DFT analysis of the CDW energetics is valid only at $T=0$, the Landau free-energy expansion allows us to also investigate the fate of the CDW transitions at finite temperatures. The close proximity between different CDW ground states at $T=0$, as shown for instance in Fig. \ref{fig:pressure}, suggests that multiple CDW transitions can take place as a function of temperature. While such a scenario was explored phenomenologically in Ref. \cite{Christensen2021} using the same free energy as in Eq.~\ref{eq:landau_tot}, our approach gives the Landau parameters of CsV$_3$Sb$_5$ directly from DFT. Experimentally, probes such as Raman spectroscopy \cite{Wu2022}, $\mu$SR \cite{yu2021evidence,khasanov2022time}, and elastoresistance \cite{Nie2022} report signatures consistent with multiple CDW transitions as temperature is changed, which may also involve time-reversal and threefold rotational symmetry-breaking.      

In order to render the problem tractable, and following the spirit of a Landau expansion, we assume that only the quadratic coefficients $\alpha_M$ and $\alpha_L$ are temperature dependent. Thus, we define {$\alpha_M\equiv \alpha_M^0(T_M-T)/T_M$} and {$\alpha_L\equiv \alpha_L^0(T_L-T)/T_L$}, and set the zero temperature quadratic coefficients $\alpha_M^0$ and $\alpha_L^0$ equal to the $\alpha_M$ and $\alpha_L$ predicted by DFT, while treating the bare critical temperatures $T_M$ and $T_L$ as free parameters. For all third and fourth order coefficients, we use the values extracted from DFT at each pressure. Because, in general, $\alpha_M^0 \neq \alpha_L^0$, the temperature phase diagrams depend on the absolute values of both $T_M$ and $T_L$, and not just on their difference. It is therefore convenient to define the average transition temperature $T_0 \equiv (T_M + T_L)/2$, which sets the temperature scale of the transition, and the dimensionless temperature difference $\delta \tau \equiv (T_M - T_L)/(T_M+T_L)$, such that $T_M=T_0(1+\delta\tau)$ and $T_L=T_0(1-\delta \tau)$. For concreteness, we consider the range $0.60<T_M/T_L<1.67$. This range is likely to include the actual value of $T_M/T_L$ given how close $\alpha_M$ and $\alpha_L$ are to each other at zero temperature.

\begin{figure}[]
   
    \includegraphics[width=1.0\linewidth]{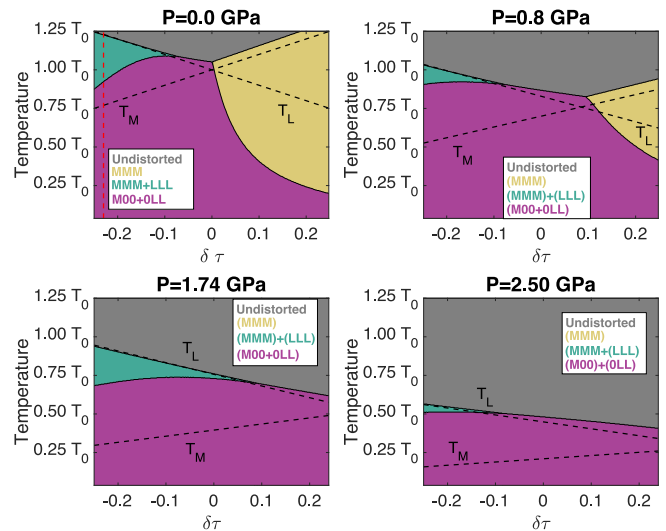}
    \caption{Finite-temperature CDW phase diagrams corresponding to different choices of the ``pure" CDW transition temperatures $T_M$ and $T_L$ (dashed black lines) at different pressures. The vertical axis corresponds to the temperature as a fraction of the average transition temperature $T_0 \equiv (T_M + T_L)/2$, whereas the horizontal axis corresponds to the relative temperature $\delta \tau \equiv (T_M - T_L)/(T_M + T_L)$. }
    \label{fig:phasediagrams}
\end{figure}

The $(\delta \tau,~T)$ CDW phase diagram is found by numerically minimizing the resulting free energy as a function of each of the six order parameter components $M_i$ and $L_i$ using the L-BFGS-G algorithm \cite{Liu1989} and the SciPy Python library \cite{Virtanen2020}, considering a grid of 300 values of $\delta \tau$ and 400 values of $T$ at each pressure.
In Fig.~\ref{fig:phasediagrams}a, we show the zero pressure CDW phase diagram. There are three different possible CDW phases, which correspond to the three lowest energy CDW structures found in DFT. While the zero temperature CDW ground state is $(M00)+(0LL)$ (staggered tri-hexagonal) for any value of $\delta \tau$, the CDW phase condensed immediately below the highest temperature transition can be either $(M00)+(0LL)$ or a different phase -- namely, $(MMM)+(LLL)$ (tri-hexagonal Star-of-David) or $MMM$ (planar tri-hexagonal). Since x-ray experiments find a CDW unit cell that is at least doubled along the $c$ axis, the $(MMM)$ phase cannot be the intermediate phase. Thus, the only scenario that gives two separate CDW transitions at zero pressure as a function of temperature, without invoking time-reversal symmetry-breaking, is a transition from an undistorted kagome lattice to the $(MMM)+(LLL)$ CDW phase, followed by a lower temperature transition to the $(M00)+(0LL)$ CDW phase. This would restrict the parameter $\delta \tau$ to $\delta \tau \lesssim -0.1$. Moreover, it would be manifested by the breaking of threefold rotational symmetry below the second CDW transition \cite{park2021electronic,Christensen2021}, in qualitative agreement with the experiments \cite{Nie2022,Xiang2021twofold}.

Following the Raman spectroscopy results of Ref.~[\onlinecite{Wu2022}], we set the ratio between the first and second transition temperatures in CsV$_3$Sb$_5$ to be $(70~K/94~K)=75\%$. Using the phase diagram at zero pressure, this sets $\delta \tau\simeq-0.23$, which implies $T_M/T_L \approx 0.63$ and is indicated by the vertical dotted red line in Figure \ref{fig:phasediagrams}. This is to be contrasted with the zero-temperature DFT result that $\alpha_M(T=0)/\alpha_L(T=0){\sim}0.87$. While such a change in the proximity between the $M$ and $L$ CDW instabilities from zero temperature to finite temperature is possible, a perhaps more likely scenario would be that one of the transitions involves the condensation of a different order parameter not captured in our DFT analysis, such as the imaginary CDW (i.e. loop-current) order parameters discussed for instance in Ref. \cite{Christensen2022}.

The DFT-extracted Landau coefficients in Table~\ref{tab:constants} can also be used to examine the behavior of the CDW phase as both temperature and pressure are changed simultaneously. As the quadratic coefficients $\alpha$ become less negative under pressure, we can expect $T_M$ and $T_L$ to decrease and the undistorted kagome phase to become stable at a wider temperature range. We use a simple parametrization of $T_M$ and $T_L$, assuming that they scale proportionally to the DFT quadratic coefficient at each finite pressure, such that
\begin{equation*}
    T_M(P,\delta \tau)=\frac{\alpha_M(P)}{\alpha_M(P=0)}T_M(P=0,\delta \tau)
\end{equation*}
and
\begin{equation*}
    T_L(P,\delta \tau)=\frac{\alpha_L(P)}{\alpha_L(P=0)}T_L(P=0,\delta \tau).
\end{equation*}
Thus, for a given value of $\Omega$ (and thus a particular combination of $T_M$ and $T_L$) at 0 GPa, we can predict the change in the CDW behavior with pressure by combining $T_M(P,\Omega)$ and $T_L(P,\Omega)$ with the finite pressure coefficients found in Table \ref{tab:constants}, as shown in Figs. \ref{fig:phasediagrams}b-d. As the pressure increases, the area associated with the $(MMM)$ phase rapidly diminishes. This is consistent with a rapidly suppressed $|\gamma_M|$ coefficient under pressure. The sequence of transitions from undistorted to $(MMM)+(LLL)$ and then to $(M00)+(0LL)$ is possible throughout the studied pressure range. The $(M00)+(0LL)$ CDW phase remains the lowest-energy low temperature structure for all pressures, due to the coefficient $\gamma_{ML}$ being significant even as the quadratic coefficients $\alpha$ approach zero.

\section{Conclusions}
\label{sec:conclusions}

In summary, we performed a detailed first principles study of the lattice energetics of the kagome metal CsV$_3$Sb$_5$ by not only considering different metastable ground state candidates, but also extracting the coefficients of the Landau free-energy expansion. Using this approach, we drew finite-temperature phase diagrams under pressure, and showed that while a scenario with two lattice transitions is not explicitly inconsistent with the DFT results, it is not favored by it. We also performed constrained structural relaxations to show that while V-only displacements of atoms lead to an energy lowering, a large portion of the energy gain is obtained through Sb displacements for all ground state candidates. This shows that even though the lattice instability is driven by V displacements, its collective nature makes the apical Sb ions essential for understanding the relative stability of different CDW phases. 

It is instructive to compare our results with experiments. Our finding that the Sb degrees of freedom give an important contribution to the energetics of the CDW phase is consistent with the conclusions of Ref. \cite{han2022orbital}, which performed x-ray absorption experiments in CsV$_3$Sb$_5$. Moreover, our finding may also be connected to the strong sensitivity of the CDW transition temperature on the $c$-axis parameter, as revealed by Ref. \cite{Qian2021}. This is because modifications of the $c$-axis lattice parameter promoted by hydrostatic or uniaxial pressure are expected to particularly impact the apical Sb ion displacements. This effect, in turn, should also affect the electronic structure by altering the overlap between the Sb $p_z$-orbitals, which are known to contribute to the Fermi surface pocket centered at the $\Gamma$ point \cite{tsirlin2022role,jeong2022crucial,Oey2022}, resulting in an interesting feedback between structural and electronic degrees of freedom.

Our analysis of the pressure dependence of the CDW phase reveals that the near-degeneracy between the $M$ and $L$ point instabilities, as reflected by their similar quadratic Landau coefficients, is lifted for high enough pressures. In particular, there is a range of pressures for which $\alpha_M$ becomes positive while $\alpha_L$ remains negative. Yet, the CDW ground state at these pressures remain the same as the one found at ambient pressure -- namely, the $(M00)+(0LL)$ staggered tri-hexagonal CDW phase. This highlights the importance of the coupling between the $M$ and $L$ CDW order parameters in promoting the CDW instability. Elucidating the microscopic origin of such a coupling will shed new light on the microscopic mechanism of the CDW instability. While a purely phononic mechanism is unlikely \cite{Christensen2021}, our results highlighting the importance of the Sb degrees of freedom show that a mechanism relying only on the van Hove singularities arising from the V orbitals may not be enough either. Taking into account the $k_z$ dispersion of the saddle points that give rise to these van Hove singularities may be important to capture the coupling between the $M$ and $L$ CDW orders. Interestingly, one of these saddle points has a strong spectral weight contribution from the Sb orbitals \cite{jeong2022crucial}. Moreover, for finite $k_z$ values, bands stemming from van Hove singularities even cross the Fermi level \cite{Oey2022}.

Previous experiments have reported evidence for a pressure-induced transition between two different CDW phases \cite{Li2022discovery,Gupta2022a}, as reflected by the double-peak structure of the superconducting dome inside the coexistence state with CDW \cite{Yu2021b, Zhang2021Pressure, Chen2021Pressurea, Chen2021Pressureb, Du2022}. Our analysis, which extends the findings first reported by us in Ref. \cite{Gupta2022a}, reveals instead that the CDW ground state remains unchanged as a function of pressure. While effects not captured by DFT may impact the small energy differences between the stable CDW states and thus favor a different ground state, it is also plausible that a distinct type of charge order is at play, such as the so-called imaginary CDW, which breaks time-reversal symmetry. 

Similarly, there is experimental evidence for two different CDW transitions as a function of temperature \cite{Wu2022,Xiang2021twofold,Nie2022}. Our finite-temperature analysis does find a narrow parameter regime in which the $(M00)+(0LL)$ staggered tri-hexagonal CDW ground state is preceded by a transition to the $(MMM)+(LLL)$ tri-hexagonal Star-of-David CDW phase. Notwithstanding the rather restrictive conditions in this parameter range, it is not clear whether this scenario could explain the experimental observations. The main signature of the $(MMM)+(LLL)$ to $(M00)+(0LL)$ transition would be the breaking of the threefold rotational symmetry of the kagome lattice. The transport data of Refs. \cite{Xiang2021twofold,Nie2022} is consistent with such a transition, as threefold rotational symmetry breaking is observed only well below the CDW transition temperature. On the other hand, the optical data of Ref. \cite{xu2022three} shows threefold rotational symmetry being broken at the same temperature as the onset of CDW. Furthermore, the Raman data of Ref. \cite{Wu2022} is consistent with two CDW transitions between structures that share the same symmetries. As with the situation of the pressure-induced CDW-to-CDW transition, it is also possible that one of the finite-temperature transitions is related instead to an imaginary charge density wave (i.e. loop currents). As discussed above, while the reported $\mu$SR data is characteristic of a time-reversal symmetry-breaking transition \cite{mielke2022time,khasanov2022time}, there is disagreement on Kerr effect data \cite{xu2022three,saykin2022_Kerr,hu2022_Kerr}. To shed more light on this issue, it would be interesting to be able to capture such loop current phases within a first-principles approach.

\acknowledgements
We thank B. Andersen and M. Christensen for fruitful discussions. ETR and TB were supported by the NSF CAREER grant DMR-2046020. RMF was supported by the Air Force Office of Scientific Research under Award No. FA9550-21-1-0423.

\providecommand{\noopsort}[1]{}\providecommand{\singleletter}[1]{#1}%

\providecommand{\noopsort}[1]{}\providecommand{\singleletter}[1]{#1}%
\begin{thebibliography}{62}%
\makeatletter
\providecommand \@ifxundefined [1]{%
 \@ifx{#1\undefined}
}%
\providecommand \@ifnum [1]{%
 \ifnum #1\expandafter \@firstoftwo
 \else \expandafter \@secondoftwo
 \fi
}%
\providecommand \@ifx [1]{%
 \ifx #1\expandafter \@firstoftwo
 \else \expandafter \@secondoftwo
 \fi
}%
\providecommand \natexlab [1]{#1}%
\providecommand \enquote  [1]{``#1''}%
\providecommand \bibnamefont  [1]{#1}%
\providecommand \bibfnamefont [1]{#1}%
\providecommand \citenamefont [1]{#1}%
\providecommand \href@noop [0]{\@secondoftwo}%
\providecommand \href [0]{\begingroup \@sanitize@url \@href}%
\providecommand \@href[1]{\@@startlink{#1}\@@href}%
\providecommand \@@href[1]{\endgroup#1\@@endlink}%
\providecommand \@sanitize@url [0]{\catcode `\\12\catcode `\$12\catcode
  `\&12\catcode `\#12\catcode `\^12\catcode `\_12\catcode `\%12\relax}%
\providecommand \@@startlink[1]{}%
\providecommand \@@endlink[0]{}%
\providecommand \url  [0]{\begingroup\@sanitize@url \@url }%
\providecommand \@url [1]{\endgroup\@href {#1}{\urlprefix }}%
\providecommand \urlprefix  [0]{URL }%
\providecommand \Eprint [0]{\href }%
\providecommand \doibase [0]{http://dx.doi.org/}%
\providecommand \selectlanguage [0]{\@gobble}%
\providecommand \bibinfo  [0]{\@secondoftwo}%
\providecommand \bibfield  [0]{\@secondoftwo}%
\providecommand \translation [1]{[#1]}%
\providecommand \BibitemOpen [0]{}%
\providecommand \bibitemStop [0]{}%
\providecommand \bibitemNoStop [0]{.\EOS\space}%
\providecommand \EOS [0]{\spacefactor3000\relax}%
\providecommand \BibitemShut  [1]{\csname bibitem#1\endcsname}%
\let\auto@bib@innerbib\@empty
\bibitem [{\citenamefont {Ortiz}\ \emph {et~al.}(2020)\citenamefont {Ortiz},
  \citenamefont {Teicher}, \citenamefont {Hu}, \citenamefont {Zuo},
  \citenamefont {Sarte}, \citenamefont {Schueller}, \citenamefont {Abeykoon},
  \citenamefont {Krogstad}, \citenamefont {Rosenkranz}, \citenamefont {Osborn}
  \emph {et~al.}}]{ortiz2020cs}%
  \BibitemOpen
  \bibfield  {author} {\bibinfo {author} {\bibfnamefont {B.~R.}\ \bibnamefont
  {Ortiz}}, \bibinfo {author} {\bibfnamefont {S.~M.}\ \bibnamefont {Teicher}},
  \bibinfo {author} {\bibfnamefont {Y.}~\bibnamefont {Hu}}, \bibinfo {author}
  {\bibfnamefont {J.~L.}\ \bibnamefont {Zuo}}, \bibinfo {author} {\bibfnamefont
  {P.~M.}\ \bibnamefont {Sarte}}, \bibinfo {author} {\bibfnamefont {E.~C.}\
  \bibnamefont {Schueller}}, \bibinfo {author} {\bibfnamefont {A.~M.}\
  \bibnamefont {Abeykoon}}, \bibinfo {author} {\bibfnamefont {M.~J.}\
  \bibnamefont {Krogstad}}, \bibinfo {author} {\bibfnamefont {S.}~\bibnamefont
  {Rosenkranz}}, \bibinfo {author} {\bibfnamefont {R.}~\bibnamefont {Osborn}},
  \emph {et~al.},\ }\href@noop {} {\bibfield  {journal} {\bibinfo  {journal}
  {Phys. Rev. Lett.}\ }\textbf {\bibinfo {volume} {125}},\ \bibinfo {pages}
  {247002} (\bibinfo {year} {2020})}\BibitemShut {NoStop}%
\bibitem [{\citenamefont {Ortiz}\ \emph
  {et~al.}(2021{\natexlab{a}})\citenamefont {Ortiz}, \citenamefont {Sarte},
  \citenamefont {Kenney}, \citenamefont {Graf}, \citenamefont {Teicher},
  \citenamefont {Seshadri},\ and\ \citenamefont
  {Wilson}}]{ortiz2021superconductivity}%
  \BibitemOpen
  \bibfield  {author} {\bibinfo {author} {\bibfnamefont {B.~R.}\ \bibnamefont
  {Ortiz}}, \bibinfo {author} {\bibfnamefont {P.~M.}\ \bibnamefont {Sarte}},
  \bibinfo {author} {\bibfnamefont {E.~M.}\ \bibnamefont {Kenney}}, \bibinfo
  {author} {\bibfnamefont {M.~J.}\ \bibnamefont {Graf}}, \bibinfo {author}
  {\bibfnamefont {S.~M.}\ \bibnamefont {Teicher}}, \bibinfo {author}
  {\bibfnamefont {R.}~\bibnamefont {Seshadri}}, \ and\ \bibinfo {author}
  {\bibfnamefont {S.~D.}\ \bibnamefont {Wilson}},\ }\href@noop {} {\bibfield
  {journal} {\bibinfo  {journal} {Phys. Rev. Mater.}\ }\textbf {\bibinfo
  {volume} {5}},\ \bibinfo {pages} {034801} (\bibinfo {year}
  {2021}{\natexlab{a}})}\BibitemShut {NoStop}%
\bibitem [{\citenamefont {Li}\ \emph {et~al.}(2021)\citenamefont {Li},
  \citenamefont {Zhang}, \citenamefont {Yilmaz}, \citenamefont {Pai},
  \citenamefont {Marvinney}, \citenamefont {Said}, \citenamefont {Yin},
  \citenamefont {Gong}, \citenamefont {Tu}, \citenamefont {Vescovo} \emph
  {et~al.}}]{li2021observation}%
  \BibitemOpen
  \bibfield  {author} {\bibinfo {author} {\bibfnamefont {H.}~\bibnamefont
  {Li}}, \bibinfo {author} {\bibfnamefont {T.}~\bibnamefont {Zhang}}, \bibinfo
  {author} {\bibfnamefont {T.}~\bibnamefont {Yilmaz}}, \bibinfo {author}
  {\bibfnamefont {Y.}~\bibnamefont {Pai}}, \bibinfo {author} {\bibfnamefont
  {C.}~\bibnamefont {Marvinney}}, \bibinfo {author} {\bibfnamefont
  {A.}~\bibnamefont {Said}}, \bibinfo {author} {\bibfnamefont {Q.}~\bibnamefont
  {Yin}}, \bibinfo {author} {\bibfnamefont {C.}~\bibnamefont {Gong}}, \bibinfo
  {author} {\bibfnamefont {Z.}~\bibnamefont {Tu}}, \bibinfo {author}
  {\bibfnamefont {E.}~\bibnamefont {Vescovo}},  \emph {et~al.},\ }\href@noop {}
  {\bibfield  {journal} {\bibinfo  {journal} {Phys. Rev. X}\ }\textbf {\bibinfo
  {volume} {11}},\ \bibinfo {pages} {031050} (\bibinfo {year}
  {2021})}\BibitemShut {NoStop}%
\bibitem [{\citenamefont {Chen}\ \emph
  {et~al.}(2021{\natexlab{a}})\citenamefont {Chen}, \citenamefont {Wang},
  \citenamefont {Yin}, \citenamefont {Gu}, \citenamefont {Jiang}, \citenamefont
  {Tu}, \citenamefont {Gong}, \citenamefont {Uwatoko}, \citenamefont {Sun},
  \citenamefont {Lei} \emph {et~al.}}]{chen2021double}%
  \BibitemOpen
  \bibfield  {author} {\bibinfo {author} {\bibfnamefont {K.}~\bibnamefont
  {Chen}}, \bibinfo {author} {\bibfnamefont {N.}~\bibnamefont {Wang}}, \bibinfo
  {author} {\bibfnamefont {Q.}~\bibnamefont {Yin}}, \bibinfo {author}
  {\bibfnamefont {Y.}~\bibnamefont {Gu}}, \bibinfo {author} {\bibfnamefont
  {K.}~\bibnamefont {Jiang}}, \bibinfo {author} {\bibfnamefont
  {Z.}~\bibnamefont {Tu}}, \bibinfo {author} {\bibfnamefont {C.}~\bibnamefont
  {Gong}}, \bibinfo {author} {\bibfnamefont {Y.}~\bibnamefont {Uwatoko}},
  \bibinfo {author} {\bibfnamefont {J.}~\bibnamefont {Sun}}, \bibinfo {author}
  {\bibfnamefont {H.}~\bibnamefont {Lei}},  \emph {et~al.},\ }\href@noop {}
  {\bibfield  {journal} {\bibinfo  {journal} {Phys. Rev. Lett.}\ }\textbf
  {\bibinfo {volume} {126}},\ \bibinfo {pages} {247001} (\bibinfo {year}
  {2021}{\natexlab{a}})}\BibitemShut {NoStop}%
\bibitem [{\citenamefont {Ortiz}\ \emph {et~al.}(2019)\citenamefont {Ortiz},
  \citenamefont {Gomes}, \citenamefont {Morey}, \citenamefont {Winiarski},
  \citenamefont {Bordelon}, \citenamefont {Mangum}, \citenamefont {Oswald},
  \citenamefont {Rodriguez-Rivera}, \citenamefont {Neilson}, \citenamefont
  {Wilson} \emph {et~al.}}]{ortiz2019new}%
  \BibitemOpen
  \bibfield  {author} {\bibinfo {author} {\bibfnamefont {B.~R.}\ \bibnamefont
  {Ortiz}}, \bibinfo {author} {\bibfnamefont {L.~C.}\ \bibnamefont {Gomes}},
  \bibinfo {author} {\bibfnamefont {J.~R.}\ \bibnamefont {Morey}}, \bibinfo
  {author} {\bibfnamefont {M.}~\bibnamefont {Winiarski}}, \bibinfo {author}
  {\bibfnamefont {M.}~\bibnamefont {Bordelon}}, \bibinfo {author}
  {\bibfnamefont {J.~S.}\ \bibnamefont {Mangum}}, \bibinfo {author}
  {\bibfnamefont {I.~W.}\ \bibnamefont {Oswald}}, \bibinfo {author}
  {\bibfnamefont {J.~A.}\ \bibnamefont {Rodriguez-Rivera}}, \bibinfo {author}
  {\bibfnamefont {J.~R.}\ \bibnamefont {Neilson}}, \bibinfo {author}
  {\bibfnamefont {S.~D.}\ \bibnamefont {Wilson}},  \emph {et~al.},\ }\href@noop
  {} {\bibfield  {journal} {\bibinfo  {journal} {Phys. Rev. Mater.}\ }\textbf
  {\bibinfo {volume} {3}},\ \bibinfo {pages} {094407} (\bibinfo {year}
  {2019})}\BibitemShut {NoStop}%
\bibitem [{\citenamefont {Ortiz}\ \emph
  {et~al.}(2021{\natexlab{b}})\citenamefont {Ortiz}, \citenamefont {Teicher},
  \citenamefont {Kautzsch}, \citenamefont {Sarte}, \citenamefont {Ratcliff},
  \citenamefont {Harter}, \citenamefont {Ruff}, \citenamefont {Seshadri},\ and\
  \citenamefont {Wilson}}]{ortiz2021fermi}%
  \BibitemOpen
  \bibfield  {author} {\bibinfo {author} {\bibfnamefont {B.~R.}\ \bibnamefont
  {Ortiz}}, \bibinfo {author} {\bibfnamefont {S.~M.}\ \bibnamefont {Teicher}},
  \bibinfo {author} {\bibfnamefont {L.}~\bibnamefont {Kautzsch}}, \bibinfo
  {author} {\bibfnamefont {P.~M.}\ \bibnamefont {Sarte}}, \bibinfo {author}
  {\bibfnamefont {N.}~\bibnamefont {Ratcliff}}, \bibinfo {author}
  {\bibfnamefont {J.}~\bibnamefont {Harter}}, \bibinfo {author} {\bibfnamefont
  {J.~P.}\ \bibnamefont {Ruff}}, \bibinfo {author} {\bibfnamefont
  {R.}~\bibnamefont {Seshadri}}, \ and\ \bibinfo {author} {\bibfnamefont
  {S.~D.}\ \bibnamefont {Wilson}},\ }\href@noop {} {\bibfield  {journal}
  {\bibinfo  {journal} {Phys. Rev. X}\ }\textbf {\bibinfo {volume} {11}},\
  \bibinfo {pages} {041030} (\bibinfo {year} {2021}{\natexlab{b}})}\BibitemShut
  {NoStop}%
\bibitem [{\citenamefont {Si}\ \emph {et~al.}(2022)\citenamefont {Si},
  \citenamefont {Lu}, \citenamefont {Sun}, \citenamefont {Liu},\ and\
  \citenamefont {Wang}}]{Si2022}%
  \BibitemOpen
  \bibfield  {author} {\bibinfo {author} {\bibfnamefont {J.-G.}\ \bibnamefont
  {Si}}, \bibinfo {author} {\bibfnamefont {W.-J.}\ \bibnamefont {Lu}}, \bibinfo
  {author} {\bibfnamefont {Y.-P.}\ \bibnamefont {Sun}}, \bibinfo {author}
  {\bibfnamefont {P.-F.}\ \bibnamefont {Liu}}, \ and\ \bibinfo {author}
  {\bibfnamefont {B.-T.}\ \bibnamefont {Wang}},\ }\href@noop {} {\bibfield
  {journal} {\bibinfo  {journal} {Phys. Rev. B}\ }\textbf {\bibinfo {volume}
  {105}},\ \bibinfo {pages} {024517} (\bibinfo {year} {2022})}\BibitemShut
  {NoStop}%
\bibitem [{\citenamefont {Zhu}\ \emph {et~al.}(2022)\citenamefont {Zhu},
  \citenamefont {Yang}, \citenamefont {Xia}, \citenamefont {Yin}, \citenamefont
  {Wang}, \citenamefont {Zhao}, \citenamefont {Dai}, \citenamefont {Tu},
  \citenamefont {Song}, \citenamefont {Tao} \emph {et~al.}}]{Zhu2022}%
  \BibitemOpen
  \bibfield  {author} {\bibinfo {author} {\bibfnamefont {C.}~\bibnamefont
  {Zhu}}, \bibinfo {author} {\bibfnamefont {X.}~\bibnamefont {Yang}}, \bibinfo
  {author} {\bibfnamefont {W.}~\bibnamefont {Xia}}, \bibinfo {author}
  {\bibfnamefont {Q.}~\bibnamefont {Yin}}, \bibinfo {author} {\bibfnamefont
  {L.}~\bibnamefont {Wang}}, \bibinfo {author} {\bibfnamefont {C.}~\bibnamefont
  {Zhao}}, \bibinfo {author} {\bibfnamefont {D.}~\bibnamefont {Dai}}, \bibinfo
  {author} {\bibfnamefont {C.}~\bibnamefont {Tu}}, \bibinfo {author}
  {\bibfnamefont {B.}~\bibnamefont {Song}}, \bibinfo {author} {\bibfnamefont
  {Z.}~\bibnamefont {Tao}},  \emph {et~al.},\ }\href@noop {} {\bibfield
  {journal} {\bibinfo  {journal} {Phys. Rev. B}\ }\textbf {\bibinfo {volume}
  {105}},\ \bibinfo {pages} {094507} (\bibinfo {year} {2022})}\BibitemShut
  {NoStop}%
\bibitem [{\citenamefont {Du}\ \emph {et~al.}(2022)\citenamefont {Du},
  \citenamefont {Li}, \citenamefont {Luo}, \citenamefont {Gong}, \citenamefont
  {Li}, \citenamefont {Jiang}, \citenamefont {Ortiz}, \citenamefont {Liu},
  \citenamefont {Xu}, \citenamefont {Wilson}, \citenamefont {Cao},
  \citenamefont {Song},\ and\ \citenamefont {Yuan}}]{Du2022}%
  \BibitemOpen
  \bibfield  {author} {\bibinfo {author} {\bibfnamefont {F.}~\bibnamefont
  {Du}}, \bibinfo {author} {\bibfnamefont {R.}~\bibnamefont {Li}}, \bibinfo
  {author} {\bibfnamefont {S.}~\bibnamefont {Luo}}, \bibinfo {author}
  {\bibfnamefont {Y.}~\bibnamefont {Gong}}, \bibinfo {author} {\bibfnamefont
  {Y.}~\bibnamefont {Li}}, \bibinfo {author} {\bibfnamefont {S.}~\bibnamefont
  {Jiang}}, \bibinfo {author} {\bibfnamefont {B.~R.}\ \bibnamefont {Ortiz}},
  \bibinfo {author} {\bibfnamefont {Y.}~\bibnamefont {Liu}}, \bibinfo {author}
  {\bibfnamefont {X.}~\bibnamefont {Xu}}, \bibinfo {author} {\bibfnamefont
  {S.~D.}\ \bibnamefont {Wilson}}, \bibinfo {author} {\bibfnamefont
  {C.}~\bibnamefont {Cao}}, \bibinfo {author} {\bibfnamefont {Y.}~\bibnamefont
  {Song}}, \ and\ \bibinfo {author} {\bibfnamefont {H.}~\bibnamefont {Yuan}},\
  }\href {\doibase 10.1103/PhysRevB.106.024516} {\bibfield  {journal} {\bibinfo
   {journal} {Phys. Rev. B}\ }\textbf {\bibinfo {volume} {106}},\ \bibinfo
  {pages} {024516} (\bibinfo {year} {2022})}\BibitemShut {NoStop}%
\bibitem [{\citenamefont {Consiglio}\ \emph {et~al.}(2022)\citenamefont
  {Consiglio}, \citenamefont {Schwemmer}, \citenamefont {Wu}, \citenamefont
  {Hanke}, \citenamefont {Neupert}, \citenamefont {Thomale}, \citenamefont
  {Sangiovanni},\ and\ \citenamefont {Di~Sante}}]{Consiglio2022}%
  \BibitemOpen
  \bibfield  {author} {\bibinfo {author} {\bibfnamefont {A.}~\bibnamefont
  {Consiglio}}, \bibinfo {author} {\bibfnamefont {T.}~\bibnamefont
  {Schwemmer}}, \bibinfo {author} {\bibfnamefont {X.}~\bibnamefont {Wu}},
  \bibinfo {author} {\bibfnamefont {W.}~\bibnamefont {Hanke}}, \bibinfo
  {author} {\bibfnamefont {T.}~\bibnamefont {Neupert}}, \bibinfo {author}
  {\bibfnamefont {R.}~\bibnamefont {Thomale}}, \bibinfo {author} {\bibfnamefont
  {G.}~\bibnamefont {Sangiovanni}}, \ and\ \bibinfo {author} {\bibfnamefont
  {D.}~\bibnamefont {Di~Sante}},\ }\href@noop {} {\bibfield  {journal}
  {\bibinfo  {journal} {Phys. Rev. B}\ }\textbf {\bibinfo {volume} {105}},\
  \bibinfo {pages} {165146} (\bibinfo {year} {2022})}\BibitemShut {NoStop}%
\bibitem [{\citenamefont {Jiang}\ \emph {et~al.}(2022)\citenamefont {Jiang},
  \citenamefont {Wu}, \citenamefont {Yin}, \citenamefont {Wang}, \citenamefont
  {Hasan}, \citenamefont {Wilson}, \citenamefont {Chen},\ and\ \citenamefont
  {Hu}}]{review1}%
  \BibitemOpen
  \bibfield  {author} {\bibinfo {author} {\bibfnamefont {K.}~\bibnamefont
  {Jiang}}, \bibinfo {author} {\bibfnamefont {T.}~\bibnamefont {Wu}}, \bibinfo
  {author} {\bibfnamefont {J.-X.}\ \bibnamefont {Yin}}, \bibinfo {author}
  {\bibfnamefont {Z.}~\bibnamefont {Wang}}, \bibinfo {author} {\bibfnamefont
  {M.~Z.}\ \bibnamefont {Hasan}}, \bibinfo {author} {\bibfnamefont {S.~D.}\
  \bibnamefont {Wilson}}, \bibinfo {author} {\bibfnamefont {X.}~\bibnamefont
  {Chen}}, \ and\ \bibinfo {author} {\bibfnamefont {J.}~\bibnamefont {Hu}},\
  }\href {\doibase 10.1093/nsr/nwac199} {\bibfield  {journal} {\bibinfo
  {journal} {Nat. Sci. Rev.}\ } (\bibinfo {year} {2022}),\
  10.1093/nsr/nwac199}\BibitemShut {NoStop}%
\bibitem [{\citenamefont {Neupert}\ \emph {et~al.}(2022)\citenamefont
  {Neupert}, \citenamefont {Denner}, \citenamefont {Yin}, \citenamefont
  {Thomale},\ and\ \citenamefont {Hasan}}]{review2}%
  \BibitemOpen
  \bibfield  {author} {\bibinfo {author} {\bibfnamefont {T.}~\bibnamefont
  {Neupert}}, \bibinfo {author} {\bibfnamefont {M.~M.}\ \bibnamefont {Denner}},
  \bibinfo {author} {\bibfnamefont {J.-X.}\ \bibnamefont {Yin}}, \bibinfo
  {author} {\bibfnamefont {R.}~\bibnamefont {Thomale}}, \ and\ \bibinfo
  {author} {\bibfnamefont {M.~Z.}\ \bibnamefont {Hasan}},\ }\href@noop {}
  {\bibfield  {journal} {\bibinfo  {journal} {Nat. Phys.}\ }\textbf {\bibinfo
  {volume} {18}},\ \bibinfo {pages} {137} (\bibinfo {year} {2022})}\BibitemShut
  {NoStop}%
\bibitem [{\citenamefont {Chen}\ \emph
  {et~al.}(2021{\natexlab{b}})\citenamefont {Chen}, \citenamefont {Zhan},
  \citenamefont {Wang}, \citenamefont {Deng}, \citenamefont {Liu},
  \citenamefont {Chen}, \citenamefont {Guo},\ and\ \citenamefont
  {Chen}}]{Chen2021Pressurea}%
  \BibitemOpen
  \bibfield  {author} {\bibinfo {author} {\bibfnamefont {X.}~\bibnamefont
  {Chen}}, \bibinfo {author} {\bibfnamefont {X.}~\bibnamefont {Zhan}}, \bibinfo
  {author} {\bibfnamefont {X.}~\bibnamefont {Wang}}, \bibinfo {author}
  {\bibfnamefont {J.}~\bibnamefont {Deng}}, \bibinfo {author} {\bibfnamefont
  {X.~B.}\ \bibnamefont {Liu}}, \bibinfo {author} {\bibfnamefont
  {X.}~\bibnamefont {Chen}}, \bibinfo {author} {\bibfnamefont {J.~G.}\
  \bibnamefont {Guo}}, \ and\ \bibinfo {author} {\bibfnamefont
  {X.}~\bibnamefont {Chen}},\ }\href {\doibase 10.1088/0256-307X/38/5/057402}
  {\bibfield  {journal} {\bibinfo  {journal} {Chinese Phys. Lett.}\ }\textbf
  {\bibinfo {volume} {38}} (\bibinfo {year} {2021}{\natexlab{b}}),\
  10.1088/0256-307X/38/5/057402}\BibitemShut {NoStop}%
\bibitem [{\citenamefont {Chen}\ \emph
  {et~al.}(2021{\natexlab{c}})\citenamefont {Chen}, \citenamefont {Wang},
  \citenamefont {Yin}, \citenamefont {Gu}, \citenamefont {Jiang}, \citenamefont
  {Tu}, \citenamefont {Gong}, \citenamefont {Uwatoko}, \citenamefont {Sun},
  \citenamefont {Lei}, \citenamefont {Hu},\ and\ \citenamefont
  {Cheng}}]{Chen2021Pressureb}%
  \BibitemOpen
  \bibfield  {author} {\bibinfo {author} {\bibfnamefont {K.~Y.}\ \bibnamefont
  {Chen}}, \bibinfo {author} {\bibfnamefont {N.~N.}\ \bibnamefont {Wang}},
  \bibinfo {author} {\bibfnamefont {Q.~W.}\ \bibnamefont {Yin}}, \bibinfo
  {author} {\bibfnamefont {Y.~H.}\ \bibnamefont {Gu}}, \bibinfo {author}
  {\bibfnamefont {K.}~\bibnamefont {Jiang}}, \bibinfo {author} {\bibfnamefont
  {Z.~J.}\ \bibnamefont {Tu}}, \bibinfo {author} {\bibfnamefont {C.~S.}\
  \bibnamefont {Gong}}, \bibinfo {author} {\bibfnamefont {Y.}~\bibnamefont
  {Uwatoko}}, \bibinfo {author} {\bibfnamefont {J.~P.}\ \bibnamefont {Sun}},
  \bibinfo {author} {\bibfnamefont {H.~C.}\ \bibnamefont {Lei}}, \bibinfo
  {author} {\bibfnamefont {J.~P.}\ \bibnamefont {Hu}}, \ and\ \bibinfo {author}
  {\bibfnamefont {J.~G.}\ \bibnamefont {Cheng}},\ }\href {\doibase
  10.1103/PhysRevLett.126.247001} {\bibfield  {journal} {\bibinfo  {journal}
  {Phys. Rev. Lett.}\ }\textbf {\bibinfo {volume} {126}} (\bibinfo {year}
  {2021}{\natexlab{c}}),\ 10.1103/PhysRevLett.126.247001}\BibitemShut {NoStop}%
\bibitem [{\citenamefont {Du}\ \emph {et~al.}(2021)\citenamefont {Du},
  \citenamefont {Luo}, \citenamefont {Ortiz}, \citenamefont {Chen},
  \citenamefont {Duan}, \citenamefont {Zhang}, \citenamefont {Lu},
  \citenamefont {Wilson}, \citenamefont {Song},\ and\ \citenamefont
  {Yuan}}]{du2021pressure}%
  \BibitemOpen
  \bibfield  {author} {\bibinfo {author} {\bibfnamefont {F.}~\bibnamefont
  {Du}}, \bibinfo {author} {\bibfnamefont {S.}~\bibnamefont {Luo}}, \bibinfo
  {author} {\bibfnamefont {B.~R.}\ \bibnamefont {Ortiz}}, \bibinfo {author}
  {\bibfnamefont {Y.}~\bibnamefont {Chen}}, \bibinfo {author} {\bibfnamefont
  {W.}~\bibnamefont {Duan}}, \bibinfo {author} {\bibfnamefont {D.}~\bibnamefont
  {Zhang}}, \bibinfo {author} {\bibfnamefont {X.}~\bibnamefont {Lu}}, \bibinfo
  {author} {\bibfnamefont {S.~D.}\ \bibnamefont {Wilson}}, \bibinfo {author}
  {\bibfnamefont {Y.}~\bibnamefont {Song}}, \ and\ \bibinfo {author}
  {\bibfnamefont {H.}~\bibnamefont {Yuan}},\ }\href@noop {} {\bibfield
  {journal} {\bibinfo  {journal} {Phys. Rev. B}\ }\textbf {\bibinfo {volume}
  {103}},\ \bibinfo {pages} {L220504} (\bibinfo {year} {2021})}\BibitemShut
  {NoStop}%
\bibitem [{\citenamefont {Qian}\ \emph {et~al.}(2021)\citenamefont {Qian},
  \citenamefont {Christensen}, \citenamefont {Hu}, \citenamefont {Saha},
  \citenamefont {Andersen}, \citenamefont {Fernandes}, \citenamefont {Birol},\
  and\ \citenamefont {Ni}}]{Qian2021}%
  \BibitemOpen
  \bibfield  {author} {\bibinfo {author} {\bibfnamefont {T.}~\bibnamefont
  {Qian}}, \bibinfo {author} {\bibfnamefont {M.~H.}\ \bibnamefont
  {Christensen}}, \bibinfo {author} {\bibfnamefont {C.}~\bibnamefont {Hu}},
  \bibinfo {author} {\bibfnamefont {A.}~\bibnamefont {Saha}}, \bibinfo {author}
  {\bibfnamefont {B.~M.}\ \bibnamefont {Andersen}}, \bibinfo {author}
  {\bibfnamefont {R.~M.}\ \bibnamefont {Fernandes}}, \bibinfo {author}
  {\bibfnamefont {T.}~\bibnamefont {Birol}}, \ and\ \bibinfo {author}
  {\bibfnamefont {N.}~\bibnamefont {Ni}},\ }\href@noop {} {\bibfield  {journal}
  {\bibinfo  {journal} {Phys. Rev. B}\ }\textbf {\bibinfo {volume} {104}},\
  \bibinfo {pages} {144506} (\bibinfo {year} {2021})}\BibitemShut {NoStop}%
\bibitem [{\citenamefont {Oey}\ \emph {et~al.}(2022)\citenamefont {Oey},
  \citenamefont {Ortiz}, \citenamefont {Kaboudvand}, \citenamefont
  {Frassineti}, \citenamefont {Garcia}, \citenamefont {Cong}, \citenamefont
  {Sanna}, \citenamefont {Mitrovi\ifmmode~\acute{c}\else \'{c}\fi{}},
  \citenamefont {Seshadri},\ and\ \citenamefont {Wilson}}]{Oey2022}%
  \BibitemOpen
  \bibfield  {author} {\bibinfo {author} {\bibfnamefont {Y.~M.}\ \bibnamefont
  {Oey}}, \bibinfo {author} {\bibfnamefont {B.~R.}\ \bibnamefont {Ortiz}},
  \bibinfo {author} {\bibfnamefont {F.}~\bibnamefont {Kaboudvand}}, \bibinfo
  {author} {\bibfnamefont {J.}~\bibnamefont {Frassineti}}, \bibinfo {author}
  {\bibfnamefont {E.}~\bibnamefont {Garcia}}, \bibinfo {author} {\bibfnamefont
  {R.}~\bibnamefont {Cong}}, \bibinfo {author} {\bibfnamefont {S.}~\bibnamefont
  {Sanna}}, \bibinfo {author} {\bibfnamefont {V.~F.}\ \bibnamefont
  {Mitrovi\ifmmode~\acute{c}\else \'{c}\fi{}}}, \bibinfo {author}
  {\bibfnamefont {R.}~\bibnamefont {Seshadri}}, \ and\ \bibinfo {author}
  {\bibfnamefont {S.~D.}\ \bibnamefont {Wilson}},\ }\href {\doibase
  10.1103/PhysRevMaterials.6.L041801} {\bibfield  {journal} {\bibinfo
  {journal} {Phys. Rev. Mater.}\ }\textbf {\bibinfo {volume} {6}},\ \bibinfo
  {pages} {L041801} (\bibinfo {year} {2022})}\BibitemShut {NoStop}%
\bibitem [{\citenamefont {Stahl}\ \emph {et~al.}(2022)\citenamefont {Stahl},
  \citenamefont {Chen}, \citenamefont {Ritschel}, \citenamefont {Shekhar},
  \citenamefont {Sadrollahi}, \citenamefont {Rahn}, \citenamefont {Ivashko},
  \citenamefont {Zimmermann}, \citenamefont {Felser},\ and\ \citenamefont
  {Geck}}]{Stahl2022}%
  \BibitemOpen
  \bibfield  {author} {\bibinfo {author} {\bibfnamefont {Q.}~\bibnamefont
  {Stahl}}, \bibinfo {author} {\bibfnamefont {D.}~\bibnamefont {Chen}},
  \bibinfo {author} {\bibfnamefont {T.}~\bibnamefont {Ritschel}}, \bibinfo
  {author} {\bibfnamefont {C.}~\bibnamefont {Shekhar}}, \bibinfo {author}
  {\bibfnamefont {E.}~\bibnamefont {Sadrollahi}}, \bibinfo {author}
  {\bibfnamefont {M.~C.}\ \bibnamefont {Rahn}}, \bibinfo {author}
  {\bibfnamefont {O.}~\bibnamefont {Ivashko}}, \bibinfo {author} {\bibfnamefont
  {M.~v.}\ \bibnamefont {Zimmermann}}, \bibinfo {author} {\bibfnamefont
  {C.}~\bibnamefont {Felser}}, \ and\ \bibinfo {author} {\bibfnamefont
  {J.}~\bibnamefont {Geck}},\ }\href {\doibase 10.1103/PhysRevB.105.195136}
  {\bibfield  {journal} {\bibinfo  {journal} {Phys. Rev. B}\ }\textbf {\bibinfo
  {volume} {105}},\ \bibinfo {pages} {195136} (\bibinfo {year}
  {2022})}\BibitemShut {NoStop}%
\bibitem [{\citenamefont {Wu}\ \emph {et~al.}(2022)\citenamefont {Wu},
  \citenamefont {Ortiz}, \citenamefont {Tan}, \citenamefont {Wilson},
  \citenamefont {Yan}, \citenamefont {Birol},\ and\ \citenamefont
  {Blumberg}}]{Wu2022}%
  \BibitemOpen
  \bibfield  {author} {\bibinfo {author} {\bibfnamefont {S.}~\bibnamefont
  {Wu}}, \bibinfo {author} {\bibfnamefont {B.~R.}\ \bibnamefont {Ortiz}},
  \bibinfo {author} {\bibfnamefont {H.}~\bibnamefont {Tan}}, \bibinfo {author}
  {\bibfnamefont {S.~D.}\ \bibnamefont {Wilson}}, \bibinfo {author}
  {\bibfnamefont {B.}~\bibnamefont {Yan}}, \bibinfo {author} {\bibfnamefont
  {T.}~\bibnamefont {Birol}}, \ and\ \bibinfo {author} {\bibfnamefont
  {G.}~\bibnamefont {Blumberg}},\ }\href@noop {} {\bibfield  {journal}
  {\bibinfo  {journal} {Phys. Rev. B}\ }\textbf {\bibinfo {volume} {105}},\
  \bibinfo {pages} {155106} (\bibinfo {year} {2022})}\BibitemShut {NoStop}%
\bibitem [{\citenamefont {Gupta}\ \emph {et~al.}(2022)\citenamefont {Gupta},
  \citenamefont {Das}, \citenamefont {Mielke}, \citenamefont {Ritz},
  \citenamefont {Hotz}, \citenamefont {Yin}, \citenamefont {Tu}, \citenamefont
  {Gong}, \citenamefont {Lei}, \citenamefont {Birol}, , \citenamefont
  {Fernandes}, \citenamefont {Guguchia}, \citenamefont {Luetkens},\ and\
  \citenamefont {Khasanov}}]{Gupta2022a}%
  \BibitemOpen
  \bibfield  {author} {\bibinfo {author} {\bibfnamefont {R.}~\bibnamefont
  {Gupta}}, \bibinfo {author} {\bibfnamefont {D.}~\bibnamefont {Das}}, \bibinfo
  {author} {\bibfnamefont {C.}~\bibnamefont {Mielke}}, \bibinfo {author}
  {\bibfnamefont {E.~T.}\ \bibnamefont {Ritz}}, \bibinfo {author}
  {\bibfnamefont {F.}~\bibnamefont {Hotz}}, \bibinfo {author} {\bibfnamefont
  {Q.}~\bibnamefont {Yin}}, \bibinfo {author} {\bibfnamefont {Z.}~\bibnamefont
  {Tu}}, \bibinfo {author} {\bibfnamefont {C.}~\bibnamefont {Gong}}, \bibinfo
  {author} {\bibfnamefont {H.}~\bibnamefont {Lei}}, \bibinfo {author}
  {\bibfnamefont {T.}~\bibnamefont {Birol}}, , \bibinfo {author} {\bibfnamefont
  {R.~M.}\ \bibnamefont {Fernandes}}, \bibinfo {author} {\bibfnamefont
  {Z.}~\bibnamefont {Guguchia}}, \bibinfo {author} {\bibfnamefont
  {H.}~\bibnamefont {Luetkens}}, \ and\ \bibinfo {author} {\bibfnamefont
  {R.}~\bibnamefont {Khasanov}},\ }\href@noop {} {\bibfield  {journal}
  {\bibinfo  {journal} {Commun. Phys.}\ }\textbf {\bibinfo {volume} {5}},\
  \bibinfo {pages} {1} (\bibinfo {year} {2022})}\BibitemShut {NoStop}%
\bibitem [{\citenamefont {Li}\ \emph {et~al.}(2022{\natexlab{a}})\citenamefont
  {Li}, \citenamefont {Fabbris}, \citenamefont {Said}, \citenamefont {Sun},
  \citenamefont {Jiang}, \citenamefont {Yin}, \citenamefont {Pai},
  \citenamefont {Yoon}, \citenamefont {Lupini}, \citenamefont {Nelson} \emph
  {et~al.}}]{Li2022discovery}%
  \BibitemOpen
  \bibfield  {author} {\bibinfo {author} {\bibfnamefont {H.}~\bibnamefont
  {Li}}, \bibinfo {author} {\bibfnamefont {G.}~\bibnamefont {Fabbris}},
  \bibinfo {author} {\bibfnamefont {A.}~\bibnamefont {Said}}, \bibinfo {author}
  {\bibfnamefont {J.}~\bibnamefont {Sun}}, \bibinfo {author} {\bibfnamefont
  {Y.-X.}\ \bibnamefont {Jiang}}, \bibinfo {author} {\bibfnamefont {J.-X.}\
  \bibnamefont {Yin}}, \bibinfo {author} {\bibfnamefont {Y.-Y.}\ \bibnamefont
  {Pai}}, \bibinfo {author} {\bibfnamefont {S.}~\bibnamefont {Yoon}}, \bibinfo
  {author} {\bibfnamefont {A.~R.}\ \bibnamefont {Lupini}}, \bibinfo {author}
  {\bibfnamefont {C.}~\bibnamefont {Nelson}},  \emph {et~al.},\ }\href@noop {}
  {\bibfield  {journal} {\bibinfo  {journal} {Nat. Commun.}\ }\textbf {\bibinfo
  {volume} {13}},\ \bibinfo {pages} {6348} (\bibinfo {year}
  {2022}{\natexlab{a}})}\BibitemShut {NoStop}%
\bibitem [{\citenamefont {Li}\ \emph {et~al.}(2022{\natexlab{b}})\citenamefont
  {Li}, \citenamefont {Wu}, \citenamefont {Liu}, \citenamefont {Polley},
  \citenamefont {Guo}, \citenamefont {Wang}, \citenamefont {Han}, \citenamefont
  {Dendzik}, \citenamefont {Berntsen}, \citenamefont {Thiagarajan} \emph
  {et~al.}}]{li2022coexistence}%
  \BibitemOpen
  \bibfield  {author} {\bibinfo {author} {\bibfnamefont {C.}~\bibnamefont
  {Li}}, \bibinfo {author} {\bibfnamefont {X.}~\bibnamefont {Wu}}, \bibinfo
  {author} {\bibfnamefont {H.}~\bibnamefont {Liu}}, \bibinfo {author}
  {\bibfnamefont {C.}~\bibnamefont {Polley}}, \bibinfo {author} {\bibfnamefont
  {Q.}~\bibnamefont {Guo}}, \bibinfo {author} {\bibfnamefont {Y.}~\bibnamefont
  {Wang}}, \bibinfo {author} {\bibfnamefont {X.}~\bibnamefont {Han}}, \bibinfo
  {author} {\bibfnamefont {M.}~\bibnamefont {Dendzik}}, \bibinfo {author}
  {\bibfnamefont {M.~H.}\ \bibnamefont {Berntsen}}, \bibinfo {author}
  {\bibfnamefont {B.}~\bibnamefont {Thiagarajan}},  \emph {et~al.},\
  }\href@noop {} {\bibfield  {journal} {\bibinfo  {journal} {Phys. Rev. Res.}\
  }\textbf {\bibinfo {volume} {4}},\ \bibinfo {pages} {033072} (\bibinfo {year}
  {2022}{\natexlab{b}})}\BibitemShut {NoStop}%
\bibitem [{\citenamefont {Kang}\ \emph
  {et~al.}(2022{\natexlab{a}})\citenamefont {Kang}, \citenamefont {Fang},
  \citenamefont {Yoo}, \citenamefont {Ortiz}, \citenamefont {Oey},
  \citenamefont {Choi}, \citenamefont {Ryu}, \citenamefont {Kim}, \citenamefont
  {Jozwiak}, \citenamefont {Bostwick} \emph {et~al.}}]{kang2022charge}%
  \BibitemOpen
  \bibfield  {author} {\bibinfo {author} {\bibfnamefont {M.}~\bibnamefont
  {Kang}}, \bibinfo {author} {\bibfnamefont {S.}~\bibnamefont {Fang}}, \bibinfo
  {author} {\bibfnamefont {J.}~\bibnamefont {Yoo}}, \bibinfo {author}
  {\bibfnamefont {B.~R.}\ \bibnamefont {Ortiz}}, \bibinfo {author}
  {\bibfnamefont {Y.~M.}\ \bibnamefont {Oey}}, \bibinfo {author} {\bibfnamefont
  {J.}~\bibnamefont {Choi}}, \bibinfo {author} {\bibfnamefont {S.~H.}\
  \bibnamefont {Ryu}}, \bibinfo {author} {\bibfnamefont {J.}~\bibnamefont
  {Kim}}, \bibinfo {author} {\bibfnamefont {C.}~\bibnamefont {Jozwiak}},
  \bibinfo {author} {\bibfnamefont {A.}~\bibnamefont {Bostwick}},  \emph
  {et~al.},\ }\href@noop {} {\bibfield  {journal} {\bibinfo  {journal} {Nat.
  Mater.}\ ,\ \bibinfo {pages} {1}} (\bibinfo {year}
  {2022}{\natexlab{a}})}\BibitemShut {NoStop}%
\bibitem [{\citenamefont {Xiang}\ \emph {et~al.}(2021)\citenamefont {Xiang},
  \citenamefont {Li}, \citenamefont {Li}, \citenamefont {Xie}, \citenamefont
  {Yang}, \citenamefont {Wang}, \citenamefont {Yao},\ and\ \citenamefont
  {Wen}}]{Xiang2021twofold}%
  \BibitemOpen
  \bibfield  {author} {\bibinfo {author} {\bibfnamefont {Y.}~\bibnamefont
  {Xiang}}, \bibinfo {author} {\bibfnamefont {Q.}~\bibnamefont {Li}}, \bibinfo
  {author} {\bibfnamefont {Y.}~\bibnamefont {Li}}, \bibinfo {author}
  {\bibfnamefont {W.}~\bibnamefont {Xie}}, \bibinfo {author} {\bibfnamefont
  {H.}~\bibnamefont {Yang}}, \bibinfo {author} {\bibfnamefont {Z.}~\bibnamefont
  {Wang}}, \bibinfo {author} {\bibfnamefont {Y.}~\bibnamefont {Yao}}, \ and\
  \bibinfo {author} {\bibfnamefont {H.-H.}\ \bibnamefont {Wen}},\ }\href@noop
  {} {\bibfield  {journal} {\bibinfo  {journal} {Nat. Commun.}\ }\textbf
  {\bibinfo {volume} {12}},\ \bibinfo {pages} {1} (\bibinfo {year}
  {2021})}\BibitemShut {NoStop}%
\bibitem [{\citenamefont {Zhao}\ \emph {et~al.}(2021)\citenamefont {Zhao},
  \citenamefont {Li}, \citenamefont {Ortiz}, \citenamefont {Teicher},
  \citenamefont {Park}, \citenamefont {Ye}, \citenamefont {Wang}, \citenamefont
  {Balents}, \citenamefont {Wilson},\ and\ \citenamefont
  {Zeljkovic}}]{zhao2021cascade}%
  \BibitemOpen
  \bibfield  {author} {\bibinfo {author} {\bibfnamefont {H.}~\bibnamefont
  {Zhao}}, \bibinfo {author} {\bibfnamefont {H.}~\bibnamefont {Li}}, \bibinfo
  {author} {\bibfnamefont {B.~R.}\ \bibnamefont {Ortiz}}, \bibinfo {author}
  {\bibfnamefont {S.~M.}\ \bibnamefont {Teicher}}, \bibinfo {author}
  {\bibfnamefont {T.}~\bibnamefont {Park}}, \bibinfo {author} {\bibfnamefont
  {M.}~\bibnamefont {Ye}}, \bibinfo {author} {\bibfnamefont {Z.}~\bibnamefont
  {Wang}}, \bibinfo {author} {\bibfnamefont {L.}~\bibnamefont {Balents}},
  \bibinfo {author} {\bibfnamefont {S.~D.}\ \bibnamefont {Wilson}}, \ and\
  \bibinfo {author} {\bibfnamefont {I.}~\bibnamefont {Zeljkovic}},\ }\href@noop
  {} {\bibfield  {journal} {\bibinfo  {journal} {Nature}\ }\textbf {\bibinfo
  {volume} {599}},\ \bibinfo {pages} {216} (\bibinfo {year}
  {2021})}\BibitemShut {NoStop}%
\bibitem [{\citenamefont {Li}\ \emph {et~al.}(2022{\natexlab{c}})\citenamefont
  {Li}, \citenamefont {Zhao}, \citenamefont {Ortiz}, \citenamefont {Park},
  \citenamefont {Ye}, \citenamefont {Balents}, \citenamefont {Wang},
  \citenamefont {Wilson},\ and\ \citenamefont {Zeljkovic}}]{Li2022rotation}%
  \BibitemOpen
  \bibfield  {author} {\bibinfo {author} {\bibfnamefont {H.}~\bibnamefont
  {Li}}, \bibinfo {author} {\bibfnamefont {H.}~\bibnamefont {Zhao}}, \bibinfo
  {author} {\bibfnamefont {B.~R.}\ \bibnamefont {Ortiz}}, \bibinfo {author}
  {\bibfnamefont {T.}~\bibnamefont {Park}}, \bibinfo {author} {\bibfnamefont
  {M.}~\bibnamefont {Ye}}, \bibinfo {author} {\bibfnamefont {L.}~\bibnamefont
  {Balents}}, \bibinfo {author} {\bibfnamefont {Z.}~\bibnamefont {Wang}},
  \bibinfo {author} {\bibfnamefont {S.~D.}\ \bibnamefont {Wilson}}, \ and\
  \bibinfo {author} {\bibfnamefont {I.}~\bibnamefont {Zeljkovic}},\ }\href@noop
  {} {\bibfield  {journal} {\bibinfo  {journal} {Nat. Phys.}\ }\textbf
  {\bibinfo {volume} {18}},\ \bibinfo {pages} {265} (\bibinfo {year}
  {2022}{\natexlab{c}})}\BibitemShut {NoStop}%
\bibitem [{\citenamefont {Nie}\ \emph {et~al.}(2022)\citenamefont {Nie},
  \citenamefont {Sun}, \citenamefont {Ma}, \citenamefont {Song}, \citenamefont
  {Zheng}, \citenamefont {Liang}, \citenamefont {Wu}, \citenamefont {Yu},
  \citenamefont {Li}, \citenamefont {Shan}, \citenamefont {Zhao}, \citenamefont
  {Li}, \citenamefont {Kang}, \citenamefont {Wu}, \citenamefont {Zhou},
  \citenamefont {Liu}, \citenamefont {Xiang}, \citenamefont {Ying},
  \citenamefont {Wang}, \citenamefont {Wu},\ and\ \citenamefont
  {Chen}}]{Nie2022}%
  \BibitemOpen
  \bibfield  {author} {\bibinfo {author} {\bibfnamefont {L.}~\bibnamefont
  {Nie}}, \bibinfo {author} {\bibfnamefont {K.}~\bibnamefont {Sun}}, \bibinfo
  {author} {\bibfnamefont {W.}~\bibnamefont {Ma}}, \bibinfo {author}
  {\bibfnamefont {D.}~\bibnamefont {Song}}, \bibinfo {author} {\bibfnamefont
  {L.}~\bibnamefont {Zheng}}, \bibinfo {author} {\bibfnamefont
  {Z.}~\bibnamefont {Liang}}, \bibinfo {author} {\bibfnamefont
  {P.}~\bibnamefont {Wu}}, \bibinfo {author} {\bibfnamefont {F.}~\bibnamefont
  {Yu}}, \bibinfo {author} {\bibfnamefont {J.}~\bibnamefont {Li}}, \bibinfo
  {author} {\bibfnamefont {M.}~\bibnamefont {Shan}}, \bibinfo {author}
  {\bibfnamefont {D.}~\bibnamefont {Zhao}}, \bibinfo {author} {\bibfnamefont
  {S.}~\bibnamefont {Li}}, \bibinfo {author} {\bibfnamefont {B.}~\bibnamefont
  {Kang}}, \bibinfo {author} {\bibfnamefont {Z.}~\bibnamefont {Wu}}, \bibinfo
  {author} {\bibfnamefont {Y.}~\bibnamefont {Zhou}}, \bibinfo {author}
  {\bibfnamefont {K.}~\bibnamefont {Liu}}, \bibinfo {author} {\bibfnamefont
  {Z.}~\bibnamefont {Xiang}}, \bibinfo {author} {\bibfnamefont
  {J.}~\bibnamefont {Ying}}, \bibinfo {author} {\bibfnamefont {Z.}~\bibnamefont
  {Wang}}, \bibinfo {author} {\bibfnamefont {T.}~\bibnamefont {Wu}}, \ and\
  \bibinfo {author} {\bibfnamefont {X.}~\bibnamefont {Chen}},\ }\href {\doibase
  10.1038/s41586-022-04493-8} {\bibfield  {journal} {\bibinfo  {journal}
  {Nature}\ }\textbf {\bibinfo {volume} {604}},\ \bibinfo {pages} {59}
  (\bibinfo {year} {2022})}\BibitemShut {NoStop}%
\bibitem [{\citenamefont {Xu}\ \emph {et~al.}(2022)\citenamefont {Xu},
  \citenamefont {Ni}, \citenamefont {Liu}, \citenamefont {Ortiz}, \citenamefont
  {Deng}, \citenamefont {Wilson}, \citenamefont {Yan}, \citenamefont
  {Balents},\ and\ \citenamefont {Wu}}]{xu2022three}%
  \BibitemOpen
  \bibfield  {author} {\bibinfo {author} {\bibfnamefont {Y.}~\bibnamefont
  {Xu}}, \bibinfo {author} {\bibfnamefont {Z.}~\bibnamefont {Ni}}, \bibinfo
  {author} {\bibfnamefont {Y.}~\bibnamefont {Liu}}, \bibinfo {author}
  {\bibfnamefont {B.~R.}\ \bibnamefont {Ortiz}}, \bibinfo {author}
  {\bibfnamefont {Q.}~\bibnamefont {Deng}}, \bibinfo {author} {\bibfnamefont
  {S.~D.}\ \bibnamefont {Wilson}}, \bibinfo {author} {\bibfnamefont
  {B.}~\bibnamefont {Yan}}, \bibinfo {author} {\bibfnamefont {L.}~\bibnamefont
  {Balents}}, \ and\ \bibinfo {author} {\bibfnamefont {L.}~\bibnamefont {Wu}},\
  }\href@noop {} {\bibfield  {journal} {\bibinfo  {journal} {Nat. Phys.}\
  }\textbf {\bibinfo {volume} {18}},\ \bibinfo {pages} {1470} (\bibinfo {year}
  {2022})}\BibitemShut {NoStop}%
\bibitem [{\citenamefont {Mielke}\ \emph {et~al.}(2022)\citenamefont {Mielke},
  \citenamefont {Das}, \citenamefont {Yin}, \citenamefont {Liu}, \citenamefont
  {Gupta}, \citenamefont {Jiang}, \citenamefont {Medarde}, \citenamefont {Wu},
  \citenamefont {Lei}, \citenamefont {Chang} \emph {et~al.}}]{mielke2022time}%
  \BibitemOpen
  \bibfield  {author} {\bibinfo {author} {\bibfnamefont {C.}~\bibnamefont
  {Mielke}}, \bibinfo {author} {\bibfnamefont {D.}~\bibnamefont {Das}},
  \bibinfo {author} {\bibfnamefont {J.-X.}\ \bibnamefont {Yin}}, \bibinfo
  {author} {\bibfnamefont {H.}~\bibnamefont {Liu}}, \bibinfo {author}
  {\bibfnamefont {R.}~\bibnamefont {Gupta}}, \bibinfo {author} {\bibfnamefont
  {Y.-X.}\ \bibnamefont {Jiang}}, \bibinfo {author} {\bibfnamefont
  {M.}~\bibnamefont {Medarde}}, \bibinfo {author} {\bibfnamefont
  {X.}~\bibnamefont {Wu}}, \bibinfo {author} {\bibfnamefont {H.}~\bibnamefont
  {Lei}}, \bibinfo {author} {\bibfnamefont {J.}~\bibnamefont {Chang}},  \emph
  {et~al.},\ }\href@noop {} {\bibfield  {journal} {\bibinfo  {journal}
  {Nature}\ }\textbf {\bibinfo {volume} {602}},\ \bibinfo {pages} {245}
  (\bibinfo {year} {2022})}\BibitemShut {NoStop}%
\bibitem [{\citenamefont {Khasanov}\ \emph {et~al.}(2022)\citenamefont
  {Khasanov}, \citenamefont {Das}, \citenamefont {Gupta}, \citenamefont
  {Mielke~III}, \citenamefont {Elender}, \citenamefont {Yin}, \citenamefont
  {Tu}, \citenamefont {Gong}, \citenamefont {Lei}, \citenamefont {Ritz} \emph
  {et~al.}}]{khasanov2022time}%
  \BibitemOpen
  \bibfield  {author} {\bibinfo {author} {\bibfnamefont {R.}~\bibnamefont
  {Khasanov}}, \bibinfo {author} {\bibfnamefont {D.}~\bibnamefont {Das}},
  \bibinfo {author} {\bibfnamefont {R.}~\bibnamefont {Gupta}}, \bibinfo
  {author} {\bibfnamefont {C.}~\bibnamefont {Mielke~III}}, \bibinfo {author}
  {\bibfnamefont {M.}~\bibnamefont {Elender}}, \bibinfo {author} {\bibfnamefont
  {Q.}~\bibnamefont {Yin}}, \bibinfo {author} {\bibfnamefont {Z.}~\bibnamefont
  {Tu}}, \bibinfo {author} {\bibfnamefont {C.}~\bibnamefont {Gong}}, \bibinfo
  {author} {\bibfnamefont {H.}~\bibnamefont {Lei}}, \bibinfo {author}
  {\bibfnamefont {E.~T.}\ \bibnamefont {Ritz}},  \emph {et~al.},\ }\href@noop
  {} {\bibfield  {journal} {\bibinfo  {journal} {Phys. Rev. Res.}\ }\textbf
  {\bibinfo {volume} {4}},\ \bibinfo {pages} {023244} (\bibinfo {year}
  {2022})}\BibitemShut {NoStop}%
\bibitem [{\citenamefont {Christensen}\ and\ \citenamefont
  {Birol}(2022)}]{christensen2022loopy}%
  \BibitemOpen
  \bibfield  {author} {\bibinfo {author} {\bibfnamefont {M.~H.}\ \bibnamefont
  {Christensen}}\ and\ \bibinfo {author} {\bibfnamefont {T.}~\bibnamefont
  {Birol}},\ }\href {\doibase 10.1038/d41586-022-00305-1} {\bibfield  {journal}
  {\bibinfo  {journal} {Nature}\ }\textbf {\bibinfo {volume} {602}},\ \bibinfo
  {pages} {216} (\bibinfo {year} {2022})}\BibitemShut {NoStop}%
\bibitem [{\citenamefont {Saykin}\ \emph {et~al.}(2022)\citenamefont {Saykin},
  \citenamefont {Farhang}, \citenamefont {Kountz}, \citenamefont {Chen},
  \citenamefont {Ortiz}, \citenamefont {Shekhar}, \citenamefont {Felser},
  \citenamefont {Wilson}, \citenamefont {Thomale}, \citenamefont {Xia} \emph
  {et~al.}}]{saykin2022_Kerr}%
  \BibitemOpen
  \bibfield  {author} {\bibinfo {author} {\bibfnamefont {D.~R.}\ \bibnamefont
  {Saykin}}, \bibinfo {author} {\bibfnamefont {C.}~\bibnamefont {Farhang}},
  \bibinfo {author} {\bibfnamefont {E.~D.}\ \bibnamefont {Kountz}}, \bibinfo
  {author} {\bibfnamefont {D.}~\bibnamefont {Chen}}, \bibinfo {author}
  {\bibfnamefont {B.~R.}\ \bibnamefont {Ortiz}}, \bibinfo {author}
  {\bibfnamefont {C.}~\bibnamefont {Shekhar}}, \bibinfo {author} {\bibfnamefont
  {C.}~\bibnamefont {Felser}}, \bibinfo {author} {\bibfnamefont {S.~D.}\
  \bibnamefont {Wilson}}, \bibinfo {author} {\bibfnamefont {R.}~\bibnamefont
  {Thomale}}, \bibinfo {author} {\bibfnamefont {J.}~\bibnamefont {Xia}},  \emph
  {et~al.},\ }\href@noop {} {\bibfield  {journal} {\bibinfo  {journal} {arXiv
  preprint arXiv:2209.10570}\ } (\bibinfo {year} {2022})}\BibitemShut {NoStop}%
\bibitem [{\citenamefont {Hu}\ \emph {et~al.}(2022)\citenamefont {Hu},
  \citenamefont {Yamane}, \citenamefont {Mattoni}, \citenamefont {Yada},
  \citenamefont {Obata}, \citenamefont {Li}, \citenamefont {Yao}, \citenamefont
  {Wang}, \citenamefont {Wang}, \citenamefont {Farhang} \emph
  {et~al.}}]{hu2022_Kerr}%
  \BibitemOpen
  \bibfield  {author} {\bibinfo {author} {\bibfnamefont {Y.}~\bibnamefont
  {Hu}}, \bibinfo {author} {\bibfnamefont {S.}~\bibnamefont {Yamane}}, \bibinfo
  {author} {\bibfnamefont {G.}~\bibnamefont {Mattoni}}, \bibinfo {author}
  {\bibfnamefont {K.}~\bibnamefont {Yada}}, \bibinfo {author} {\bibfnamefont
  {K.}~\bibnamefont {Obata}}, \bibinfo {author} {\bibfnamefont
  {Y.}~\bibnamefont {Li}}, \bibinfo {author} {\bibfnamefont {Y.}~\bibnamefont
  {Yao}}, \bibinfo {author} {\bibfnamefont {Z.}~\bibnamefont {Wang}}, \bibinfo
  {author} {\bibfnamefont {J.}~\bibnamefont {Wang}}, \bibinfo {author}
  {\bibfnamefont {C.}~\bibnamefont {Farhang}},  \emph {et~al.},\ }\href@noop {}
  {\bibfield  {journal} {\bibinfo  {journal} {arXiv preprint arXiv:2208.08036}\
  } (\bibinfo {year} {2022})}\BibitemShut {NoStop}%
\bibitem [{\citenamefont {Kang}\ \emph
  {et~al.}(2022{\natexlab{b}})\citenamefont {Kang}, \citenamefont {Fang},
  \citenamefont {Kim}, \citenamefont {Ortiz}, \citenamefont {Ryu},
  \citenamefont {Kim}, \citenamefont {Yoo}, \citenamefont {Sangiovanni},
  \citenamefont {Di~Sante}, \citenamefont {Park} \emph
  {et~al.}}]{kang2022twofold}%
  \BibitemOpen
  \bibfield  {author} {\bibinfo {author} {\bibfnamefont {M.}~\bibnamefont
  {Kang}}, \bibinfo {author} {\bibfnamefont {S.}~\bibnamefont {Fang}}, \bibinfo
  {author} {\bibfnamefont {J.-K.}\ \bibnamefont {Kim}}, \bibinfo {author}
  {\bibfnamefont {B.~R.}\ \bibnamefont {Ortiz}}, \bibinfo {author}
  {\bibfnamefont {S.~H.}\ \bibnamefont {Ryu}}, \bibinfo {author} {\bibfnamefont
  {J.}~\bibnamefont {Kim}}, \bibinfo {author} {\bibfnamefont {J.}~\bibnamefont
  {Yoo}}, \bibinfo {author} {\bibfnamefont {G.}~\bibnamefont {Sangiovanni}},
  \bibinfo {author} {\bibfnamefont {D.}~\bibnamefont {Di~Sante}}, \bibinfo
  {author} {\bibfnamefont {B.-G.}\ \bibnamefont {Park}},  \emph {et~al.},\
  }\href@noop {} {\bibfield  {journal} {\bibinfo  {journal} {Nat. Phys.}\
  }\textbf {\bibinfo {volume} {18}},\ \bibinfo {pages} {301} (\bibinfo {year}
  {2022}{\natexlab{b}})}\BibitemShut {NoStop}%
\bibitem [{\citenamefont {Denner}\ \emph {et~al.}(2021)\citenamefont {Denner},
  \citenamefont {Thomale},\ and\ \citenamefont {Neupert}}]{Denner2021}%
  \BibitemOpen
  \bibfield  {author} {\bibinfo {author} {\bibfnamefont {M.~M.}\ \bibnamefont
  {Denner}}, \bibinfo {author} {\bibfnamefont {R.}~\bibnamefont {Thomale}}, \
  and\ \bibinfo {author} {\bibfnamefont {T.}~\bibnamefont {Neupert}},\ }\href
  {\doibase 10.1103/PhysRevLett.127.217601} {\bibfield  {journal} {\bibinfo
  {journal} {Phys. Rev. Lett.}\ }\textbf {\bibinfo {volume} {127}},\ \bibinfo
  {pages} {217601} (\bibinfo {year} {2021})}\BibitemShut {NoStop}%
\bibitem [{\citenamefont {Park}\ \emph {et~al.}(2021)\citenamefont {Park},
  \citenamefont {Ye},\ and\ \citenamefont {Balents}}]{park2021electronic}%
  \BibitemOpen
  \bibfield  {author} {\bibinfo {author} {\bibfnamefont {T.}~\bibnamefont
  {Park}}, \bibinfo {author} {\bibfnamefont {M.}~\bibnamefont {Ye}}, \ and\
  \bibinfo {author} {\bibfnamefont {L.}~\bibnamefont {Balents}},\ }\href@noop
  {} {\bibfield  {journal} {\bibinfo  {journal} {Phys. Rev. B}\ }\textbf
  {\bibinfo {volume} {104}},\ \bibinfo {pages} {035142} (\bibinfo {year}
  {2021})}\BibitemShut {NoStop}%
\bibitem [{\citenamefont {Lin}\ and\ \citenamefont
  {Nandkishore}(2021)}]{Nandkishore2021}%
  \BibitemOpen
  \bibfield  {author} {\bibinfo {author} {\bibfnamefont {Y.-P.}\ \bibnamefont
  {Lin}}\ and\ \bibinfo {author} {\bibfnamefont {R.~M.}\ \bibnamefont
  {Nandkishore}},\ }\href {\doibase 10.1103/PhysRevB.104.045122} {\bibfield
  {journal} {\bibinfo  {journal} {Phys. Rev. B}\ }\textbf {\bibinfo {volume}
  {104}},\ \bibinfo {pages} {045122} (\bibinfo {year} {2021})}\BibitemShut
  {NoStop}%
\bibitem [{\citenamefont {Christensen}\ \emph {et~al.}(2021)\citenamefont
  {Christensen}, \citenamefont {Birol}, \citenamefont {Andersen},\ and\
  \citenamefont {Fernandes}}]{Christensen2021}%
  \BibitemOpen
  \bibfield  {author} {\bibinfo {author} {\bibfnamefont {M.~H.}\ \bibnamefont
  {Christensen}}, \bibinfo {author} {\bibfnamefont {T.}~\bibnamefont {Birol}},
  \bibinfo {author} {\bibfnamefont {B.~M.}\ \bibnamefont {Andersen}}, \ and\
  \bibinfo {author} {\bibfnamefont {R.~M.}\ \bibnamefont {Fernandes}},\
  }\href@noop {} {\bibfield  {journal} {\bibinfo  {journal} {Phys. Rev. B}\
  }\textbf {\bibinfo {volume} {104}},\ \bibinfo {pages} {214513} (\bibinfo
  {year} {2021})}\BibitemShut {NoStop}%
\bibitem [{\citenamefont {Tazai}\ \emph {et~al.}(2022)\citenamefont {Tazai},
  \citenamefont {Yamakawa}, \citenamefont {Onari},\ and\ \citenamefont
  {Kontani}}]{Kontani2022}%
  \BibitemOpen
  \bibfield  {author} {\bibinfo {author} {\bibfnamefont {R.}~\bibnamefont
  {Tazai}}, \bibinfo {author} {\bibfnamefont {Y.}~\bibnamefont {Yamakawa}},
  \bibinfo {author} {\bibfnamefont {S.}~\bibnamefont {Onari}}, \ and\ \bibinfo
  {author} {\bibfnamefont {H.}~\bibnamefont {Kontani}},\ }\href@noop {}
  {\bibfield  {journal} {\bibinfo  {journal} {Sci. Adv.}\ }\textbf {\bibinfo
  {volume} {8}},\ \bibinfo {pages} {eabl4108} (\bibinfo {year}
  {2022})}\BibitemShut {NoStop}%
\bibitem [{\citenamefont {Ratcliff}\ \emph {et~al.}(2021)\citenamefont
  {Ratcliff}, \citenamefont {Hallett}, \citenamefont {Ortiz}, \citenamefont
  {Wilson},\ and\ \citenamefont {Harter}}]{Ratcliff2021}%
  \BibitemOpen
  \bibfield  {author} {\bibinfo {author} {\bibfnamefont {N.}~\bibnamefont
  {Ratcliff}}, \bibinfo {author} {\bibfnamefont {L.}~\bibnamefont {Hallett}},
  \bibinfo {author} {\bibfnamefont {B.~R.}\ \bibnamefont {Ortiz}}, \bibinfo
  {author} {\bibfnamefont {S.~D.}\ \bibnamefont {Wilson}}, \ and\ \bibinfo
  {author} {\bibfnamefont {J.~W.}\ \bibnamefont {Harter}},\ }\href@noop {}
  {\bibfield  {journal} {\bibinfo  {journal} {Phys. Rev. Mater.}\ }\textbf
  {\bibinfo {volume} {5}},\ \bibinfo {pages} {L111801} (\bibinfo {year}
  {2021})}\BibitemShut {NoStop}%
\bibitem [{\citenamefont {Tan}\ \emph {et~al.}(2021)\citenamefont {Tan},
  \citenamefont {Liu}, \citenamefont {Wang},\ and\ \citenamefont
  {Yan}}]{Binghai2021}%
  \BibitemOpen
  \bibfield  {author} {\bibinfo {author} {\bibfnamefont {H.}~\bibnamefont
  {Tan}}, \bibinfo {author} {\bibfnamefont {Y.}~\bibnamefont {Liu}}, \bibinfo
  {author} {\bibfnamefont {Z.}~\bibnamefont {Wang}}, \ and\ \bibinfo {author}
  {\bibfnamefont {B.}~\bibnamefont {Yan}},\ }\href {\doibase
  10.1103/PhysRevLett.127.046401} {\bibfield  {journal} {\bibinfo  {journal}
  {Phys. Rev. Lett.}\ }\textbf {\bibinfo {volume} {127}},\ \bibinfo {pages}
  {046401} (\bibinfo {year} {2021})}\BibitemShut {NoStop}%
\bibitem [{\citenamefont {Subedi}(2022)}]{Subedi2022}%
  \BibitemOpen
  \bibfield  {author} {\bibinfo {author} {\bibfnamefont {A.}~\bibnamefont
  {Subedi}},\ }\href {\doibase 10.1103/PhysRevMaterials.6.015001} {\bibfield
  {journal} {\bibinfo  {journal} {Phys. Rev. Mater.}\ }\textbf {\bibinfo
  {volume} {6}},\ \bibinfo {pages} {015001} (\bibinfo {year}
  {2022})}\BibitemShut {NoStop}%
\bibitem [{\citenamefont {Tsirlin}\ \emph {et~al.}(2022)\citenamefont
  {Tsirlin}, \citenamefont {Fertey}, \citenamefont {Ortiz}, \citenamefont
  {Klis}, \citenamefont {Merkl}, \citenamefont {Dressel}, \citenamefont
  {Wilson},\ and\ \citenamefont {Uykur}}]{tsirlin2022role}%
  \BibitemOpen
  \bibfield  {author} {\bibinfo {author} {\bibfnamefont {A.}~\bibnamefont
  {Tsirlin}}, \bibinfo {author} {\bibfnamefont {P.}~\bibnamefont {Fertey}},
  \bibinfo {author} {\bibfnamefont {B.~R.}\ \bibnamefont {Ortiz}}, \bibinfo
  {author} {\bibfnamefont {B.}~\bibnamefont {Klis}}, \bibinfo {author}
  {\bibfnamefont {V.}~\bibnamefont {Merkl}}, \bibinfo {author} {\bibfnamefont
  {M.}~\bibnamefont {Dressel}}, \bibinfo {author} {\bibfnamefont
  {S.}~\bibnamefont {Wilson}}, \ and\ \bibinfo {author} {\bibfnamefont
  {E.}~\bibnamefont {Uykur}},\ }\href@noop {} {\bibfield  {journal} {\bibinfo
  {journal} {SciPost Phys.}\ }\textbf {\bibinfo {volume} {12}},\ \bibinfo
  {pages} {049} (\bibinfo {year} {2022})}\BibitemShut {NoStop}%
\bibitem [{\citenamefont {Jiang}\ \emph {et~al.}(2021)\citenamefont {Jiang},
  \citenamefont {Yin}, \citenamefont {Denner}, \citenamefont {Shumiya},
  \citenamefont {Ortiz}, \citenamefont {Xu}, \citenamefont {Guguchia},
  \citenamefont {He}, \citenamefont {Hossain}, \citenamefont {Liu} \emph
  {et~al.}}]{hasan2021}%
  \BibitemOpen
  \bibfield  {author} {\bibinfo {author} {\bibfnamefont {Y.-X.}\ \bibnamefont
  {Jiang}}, \bibinfo {author} {\bibfnamefont {J.-X.}\ \bibnamefont {Yin}},
  \bibinfo {author} {\bibfnamefont {M.~M.}\ \bibnamefont {Denner}}, \bibinfo
  {author} {\bibfnamefont {N.}~\bibnamefont {Shumiya}}, \bibinfo {author}
  {\bibfnamefont {B.~R.}\ \bibnamefont {Ortiz}}, \bibinfo {author}
  {\bibfnamefont {G.}~\bibnamefont {Xu}}, \bibinfo {author} {\bibfnamefont
  {Z.}~\bibnamefont {Guguchia}}, \bibinfo {author} {\bibfnamefont
  {J.}~\bibnamefont {He}}, \bibinfo {author} {\bibfnamefont {M.~S.}\
  \bibnamefont {Hossain}}, \bibinfo {author} {\bibfnamefont {X.}~\bibnamefont
  {Liu}},  \emph {et~al.},\ }\href@noop {} {\bibfield  {journal} {\bibinfo
  {journal} {Nat. Mater.}\ }\textbf {\bibinfo {volume} {20}},\ \bibinfo {pages}
  {1353} (\bibinfo {year} {2021})}\BibitemShut {NoStop}%
\bibitem [{\citenamefont {Feng}\ \emph {et~al.}(2021)\citenamefont {Feng},
  \citenamefont {Jiang}, \citenamefont {Wang},\ and\ \citenamefont
  {Hu}}]{feng2021chiral}%
  \BibitemOpen
  \bibfield  {author} {\bibinfo {author} {\bibfnamefont {X.}~\bibnamefont
  {Feng}}, \bibinfo {author} {\bibfnamefont {K.}~\bibnamefont {Jiang}},
  \bibinfo {author} {\bibfnamefont {Z.}~\bibnamefont {Wang}}, \ and\ \bibinfo
  {author} {\bibfnamefont {J.}~\bibnamefont {Hu}},\ }\href@noop {} {\bibfield
  {journal} {\bibinfo  {journal} {Sci. Bull.}\ }\textbf {\bibinfo {volume}
  {66}},\ \bibinfo {pages} {1384} (\bibinfo {year} {2021})}\BibitemShut
  {NoStop}%
\bibitem [{\citenamefont {Christensen}\ \emph {et~al.}(2022)\citenamefont
  {Christensen}, \citenamefont {Birol}, \citenamefont {Andersen},\ and\
  \citenamefont {Fernandes}}]{Christensen2022}%
  \BibitemOpen
  \bibfield  {author} {\bibinfo {author} {\bibfnamefont {M.~H.}\ \bibnamefont
  {Christensen}}, \bibinfo {author} {\bibfnamefont {T.}~\bibnamefont {Birol}},
  \bibinfo {author} {\bibfnamefont {B.~M.}\ \bibnamefont {Andersen}}, \ and\
  \bibinfo {author} {\bibfnamefont {R.~M.}\ \bibnamefont {Fernandes}},\
  }\href@noop {} {\bibfield  {journal} {\bibinfo  {journal} {Phys. Rev. B}\
  }\textbf {\bibinfo {volume} {106}},\ \bibinfo {pages} {144504} (\bibinfo
  {year} {2022})}\BibitemShut {NoStop}%
\bibitem [{\citenamefont {Jeong}\ \emph {et~al.}(2022)\citenamefont {Jeong},
  \citenamefont {Yang}, \citenamefont {Kim}, \citenamefont {Kim}, \citenamefont
  {Lee},\ and\ \citenamefont {Han}}]{jeong2022crucial}%
  \BibitemOpen
  \bibfield  {author} {\bibinfo {author} {\bibfnamefont {M.~Y.}\ \bibnamefont
  {Jeong}}, \bibinfo {author} {\bibfnamefont {H.-J.}\ \bibnamefont {Yang}},
  \bibinfo {author} {\bibfnamefont {H.~S.}\ \bibnamefont {Kim}}, \bibinfo
  {author} {\bibfnamefont {Y.~B.}\ \bibnamefont {Kim}}, \bibinfo {author}
  {\bibfnamefont {S.}~\bibnamefont {Lee}}, \ and\ \bibinfo {author}
  {\bibfnamefont {M.~J.}\ \bibnamefont {Han}},\ }\href {\doibase
  10.1103/PhysRevB.105.235145} {\bibfield  {journal} {\bibinfo  {journal}
  {Phys. Rev. B}\ }\textbf {\bibinfo {volume} {105}},\ \bibinfo {pages}
  {235145} (\bibinfo {year} {2022})}\BibitemShut {NoStop}%
\bibitem [{\citenamefont {Han}\ \emph {et~al.}(2022)\citenamefont {Han},
  \citenamefont {Tang}, \citenamefont {Li}, \citenamefont {Liu}, \citenamefont
  {Liu}, \citenamefont {Gou}, \citenamefont {Wu}, \citenamefont {Zhou},
  \citenamefont {Yang}, \citenamefont {Diao} \emph {et~al.}}]{han2022orbital}%
  \BibitemOpen
  \bibfield  {author} {\bibinfo {author} {\bibfnamefont {S.}~\bibnamefont
  {Han}}, \bibinfo {author} {\bibfnamefont {C.~S.}\ \bibnamefont {Tang}},
  \bibinfo {author} {\bibfnamefont {L.}~\bibnamefont {Li}}, \bibinfo {author}
  {\bibfnamefont {Y.}~\bibnamefont {Liu}}, \bibinfo {author} {\bibfnamefont
  {H.}~\bibnamefont {Liu}}, \bibinfo {author} {\bibfnamefont {J.}~\bibnamefont
  {Gou}}, \bibinfo {author} {\bibfnamefont {J.}~\bibnamefont {Wu}}, \bibinfo
  {author} {\bibfnamefont {D.}~\bibnamefont {Zhou}}, \bibinfo {author}
  {\bibfnamefont {P.}~\bibnamefont {Yang}}, \bibinfo {author} {\bibfnamefont
  {C.}~\bibnamefont {Diao}},  \emph {et~al.},\ }\href@noop {} {\bibfield
  {journal} {\bibinfo  {journal} {Adv. Mater.}\ ,\ \bibinfo {pages} {2209010}}
  (\bibinfo {year} {2022})}\BibitemShut {NoStop}%
\bibitem [{\citenamefont {Guguchia}\ \emph {et~al.}(2022)\citenamefont
  {Guguchia}, \citenamefont {Mielke~III}, \citenamefont {Das}, \citenamefont
  {Gupta}, \citenamefont {Yin}, \citenamefont {Liu}, \citenamefont {Yin},
  \citenamefont {Christensen}, \citenamefont {Tu}, \citenamefont {Gong} \emph
  {et~al.}}]{guguchia2022tunable}%
  \BibitemOpen
  \bibfield  {author} {\bibinfo {author} {\bibfnamefont {Z.}~\bibnamefont
  {Guguchia}}, \bibinfo {author} {\bibfnamefont {C.}~\bibnamefont
  {Mielke~III}}, \bibinfo {author} {\bibfnamefont {D.}~\bibnamefont {Das}},
  \bibinfo {author} {\bibfnamefont {R.}~\bibnamefont {Gupta}}, \bibinfo
  {author} {\bibfnamefont {J.-X.}\ \bibnamefont {Yin}}, \bibinfo {author}
  {\bibfnamefont {H.}~\bibnamefont {Liu}}, \bibinfo {author} {\bibfnamefont
  {Q.}~\bibnamefont {Yin}}, \bibinfo {author} {\bibfnamefont {M.}~\bibnamefont
  {Christensen}}, \bibinfo {author} {\bibfnamefont {Z.}~\bibnamefont {Tu}},
  \bibinfo {author} {\bibfnamefont {C.}~\bibnamefont {Gong}},  \emph {et~al.},\
  }\href@noop {} {\bibfield  {journal} {\bibinfo  {journal} {arXiv preprint
  arXiv:2202.07713}\ } (\bibinfo {year} {2022})}\BibitemShut {NoStop}%
\bibitem [{\citenamefont {Kresse}\ and\ \citenamefont
  {Hafner}(1993)}]{kresse1993ab}%
  \BibitemOpen
  \bibfield  {author} {\bibinfo {author} {\bibfnamefont {G.}~\bibnamefont
  {Kresse}}\ and\ \bibinfo {author} {\bibfnamefont {J.}~\bibnamefont
  {Hafner}},\ }\href@noop {} {\bibfield  {journal} {\bibinfo  {journal} {Phys.
  Rev. B}\ }\textbf {\bibinfo {volume} {47}},\ \bibinfo {pages} {558} (\bibinfo
  {year} {1993})}\BibitemShut {NoStop}%
\bibitem [{\citenamefont {Kresse}\ and\ \citenamefont
  {Furthm{\"u}ller}(1996{\natexlab{a}})}]{kresse1996efficiency}%
  \BibitemOpen
  \bibfield  {author} {\bibinfo {author} {\bibfnamefont {G.}~\bibnamefont
  {Kresse}}\ and\ \bibinfo {author} {\bibfnamefont {J.}~\bibnamefont
  {Furthm{\"u}ller}},\ }\href@noop {} {\bibfield  {journal} {\bibinfo
  {journal} {Comput. Mater. Sci.}\ }\textbf {\bibinfo {volume} {6}},\ \bibinfo
  {pages} {15} (\bibinfo {year} {1996}{\natexlab{a}})}\BibitemShut {NoStop}%
\bibitem [{\citenamefont {Kresse}\ and\ \citenamefont
  {Furthm{\"u}ller}(1996{\natexlab{b}})}]{kresse1996efficient}%
  \BibitemOpen
  \bibfield  {author} {\bibinfo {author} {\bibfnamefont {G.}~\bibnamefont
  {Kresse}}\ and\ \bibinfo {author} {\bibfnamefont {J.}~\bibnamefont
  {Furthm{\"u}ller}},\ }\href@noop {} {\bibfield  {journal} {\bibinfo
  {journal} {Phys. Rev. B}\ }\textbf {\bibinfo {volume} {54}},\ \bibinfo
  {pages} {11169} (\bibinfo {year} {1996}{\natexlab{b}})}\BibitemShut {NoStop}%
\bibitem [{\citenamefont {Cho}\ \emph {et~al.}(2021)\citenamefont {Cho},
  \citenamefont {Ma}, \citenamefont {Xia}, \citenamefont {Yang}, \citenamefont
  {Liu}, \citenamefont {Huang}, \citenamefont {Jiang}, \citenamefont {Lu},
  \citenamefont {Liu}, \citenamefont {Liu} \emph {et~al.}}]{cho2021emergence}%
  \BibitemOpen
  \bibfield  {author} {\bibinfo {author} {\bibfnamefont {S.}~\bibnamefont
  {Cho}}, \bibinfo {author} {\bibfnamefont {H.}~\bibnamefont {Ma}}, \bibinfo
  {author} {\bibfnamefont {W.}~\bibnamefont {Xia}}, \bibinfo {author}
  {\bibfnamefont {Y.}~\bibnamefont {Yang}}, \bibinfo {author} {\bibfnamefont
  {Z.}~\bibnamefont {Liu}}, \bibinfo {author} {\bibfnamefont {Z.}~\bibnamefont
  {Huang}}, \bibinfo {author} {\bibfnamefont {Z.}~\bibnamefont {Jiang}},
  \bibinfo {author} {\bibfnamefont {X.}~\bibnamefont {Lu}}, \bibinfo {author}
  {\bibfnamefont {J.}~\bibnamefont {Liu}}, \bibinfo {author} {\bibfnamefont
  {Z.}~\bibnamefont {Liu}},  \emph {et~al.},\ }\href@noop {} {\bibfield
  {journal} {\bibinfo  {journal} {Phys. Rev. Lett.}\ }\textbf {\bibinfo
  {volume} {127}},\ \bibinfo {pages} {236401} (\bibinfo {year}
  {2021})}\BibitemShut {NoStop}%
\bibitem [{Note1()}]{Note1}%
  \BibitemOpen
  \bibinfo {note} {We use the word `trilinear' to refer to expressions which
  involve three different order parameter components to differentiate them from
  the more common cubic terms such as $M^3$. This is a more general use the
  term than, for example, in the ferroelectrics community where the word
  `trilinear' is used to refer to a combination of three different irreps
  (${\propto }P_1 Q_1 R_3$, etc.). For example, see Ref.~[\protect
  \rev@citealpnum {Etxebarria2010, Mulder2013, Li2020Barrier}].}\BibitemShut
  {Stop}%
\bibitem [{Note2()}]{Note2}%
  \BibitemOpen
  \bibinfo {note} {All numerical fits were performed using the SciPy Python
  library. In order to insure a good fit to the 11 coefficients in Equations
  \ref {eq:landau_M}-\ref {eq:landau_mixed}, the energies associated with
  distortions in 13 directions in parameter space were calculated, then
  simultaneously fit using the non-linear least-squares method. These
  directions were selected to insure that the resulting system of least-squares
  normal equations spanned the space of all unknown coefficients, and consist
  of $(L00)$, $(LL0)$, $(M00)+(0LL)$, $(M00)+(LL0)$, $(MM0)$, $(M00)$,
  $(MM0)+(LL0)$, $(MMM)$, $(MMM)+(L00)$, $(MMM)+(LLL)$, $(M00)+(\protect \frac
  {L}{2}LL)$, and $(\protect \frac {M}{2}00)+(LLL)$. A Landau free energy
  including terms up to 5$^{th}$ order showed no appreciable change in the
  2$^{nd}$- through 4$^{th}$-order coefficients. Note that for the 3.50 GPa
  pressure data, the M-point phonons are stable ($\alpha _M>0$), but just
  barely so. This makes the energy surface highly sensitive to distortions of
  pure $M_1^+$ order, and leads to difficulties in distinguishing between
  even-order terms in the energy expansion that involve $M_1^+$ distortions
  only. Thus, the quantitative accuracy of $\alpha _M$, $u_M$, and $\lambda _M$
  at this pressure value may be less robust than for other values in Table \ref
  {tab:constants}}\BibitemShut {NoStop}%
\bibitem [{sto()}]{stokes_iso}%
  \BibitemOpen
  \href@noop {} {}\bibinfo {note} {H. T. Stokes, D. M. Hatch, and B. J.
  Campbell, ISODISTORT, ISOTROPY Software Suite, iso.byu.edu.}\BibitemShut
  {Stop}%
\bibitem [{\citenamefont {Stokes}\ \emph {et~al.}(2006)\citenamefont {Stokes},
  \citenamefont {Hatch}, \citenamefont {Campbell},\ and\ \citenamefont
  {Tanner}}]{stokes2006isodisplace}%
  \BibitemOpen
  \bibfield  {author} {\bibinfo {author} {\bibfnamefont {H.~T.}\ \bibnamefont
  {Stokes}}, \bibinfo {author} {\bibfnamefont {D.~M.}\ \bibnamefont {Hatch}},
  \bibinfo {author} {\bibfnamefont {B.~J.}\ \bibnamefont {Campbell}}, \ and\
  \bibinfo {author} {\bibfnamefont {D.~E.}\ \bibnamefont {Tanner}},\
  }\href@noop {} {\bibfield  {journal} {\bibinfo  {journal} {J. Appl.
  Crystallogr.}\ }\textbf {\bibinfo {volume} {39}},\ \bibinfo {pages} {607}
  (\bibinfo {year} {2006})}\BibitemShut {NoStop}%
\bibitem [{\citenamefont {Yu}\ \emph {et~al.}(2021{\natexlab{a}})\citenamefont
  {Yu}, \citenamefont {Ma}, \citenamefont {Zhuo}, \citenamefont {Liu},
  \citenamefont {Wen}, \citenamefont {Lei}, \citenamefont {Ying},\ and\
  \citenamefont {Chen}}]{Yu2021b}%
  \BibitemOpen
  \bibfield  {author} {\bibinfo {author} {\bibfnamefont {F.~H.}\ \bibnamefont
  {Yu}}, \bibinfo {author} {\bibfnamefont {D.~H.}\ \bibnamefont {Ma}}, \bibinfo
  {author} {\bibfnamefont {W.~Z.}\ \bibnamefont {Zhuo}}, \bibinfo {author}
  {\bibfnamefont {S.~Q.}\ \bibnamefont {Liu}}, \bibinfo {author} {\bibfnamefont
  {X.~K.}\ \bibnamefont {Wen}}, \bibinfo {author} {\bibfnamefont
  {B.}~\bibnamefont {Lei}}, \bibinfo {author} {\bibfnamefont {J.~J.}\
  \bibnamefont {Ying}}, \ and\ \bibinfo {author} {\bibfnamefont {X.~H.}\
  \bibnamefont {Chen}},\ }\href {\doibase 10.1038/s41467-021-23928-w}
  {\bibfield  {journal} {\bibinfo  {journal} {Nat. Commun.}\ }\textbf {\bibinfo
  {volume} {12}} (\bibinfo {year} {2021}{\natexlab{a}}),\
  10.1038/s41467-021-23928-w}\BibitemShut {NoStop}%
\bibitem [{\citenamefont {Zhang}\ \emph {et~al.}(2021)\citenamefont {Zhang},
  \citenamefont {Chen}, \citenamefont {Zhou}, \citenamefont {Yuan},
  \citenamefont {Wang}, \citenamefont {Wang}, \citenamefont {Yang},
  \citenamefont {An}, \citenamefont {Zhang}, \citenamefont {Zhu}, \citenamefont
  {Zhou}, \citenamefont {Chen}, \citenamefont {Zhou},\ and\ \citenamefont
  {Yang}}]{Zhang2021Pressure}%
  \BibitemOpen
  \bibfield  {author} {\bibinfo {author} {\bibfnamefont {Z.}~\bibnamefont
  {Zhang}}, \bibinfo {author} {\bibfnamefont {Z.}~\bibnamefont {Chen}},
  \bibinfo {author} {\bibfnamefont {Y.}~\bibnamefont {Zhou}}, \bibinfo {author}
  {\bibfnamefont {Y.}~\bibnamefont {Yuan}}, \bibinfo {author} {\bibfnamefont
  {S.}~\bibnamefont {Wang}}, \bibinfo {author} {\bibfnamefont {J.}~\bibnamefont
  {Wang}}, \bibinfo {author} {\bibfnamefont {H.}~\bibnamefont {Yang}}, \bibinfo
  {author} {\bibfnamefont {C.}~\bibnamefont {An}}, \bibinfo {author}
  {\bibfnamefont {L.}~\bibnamefont {Zhang}}, \bibinfo {author} {\bibfnamefont
  {X.}~\bibnamefont {Zhu}}, \bibinfo {author} {\bibfnamefont {Y.}~\bibnamefont
  {Zhou}}, \bibinfo {author} {\bibfnamefont {X.}~\bibnamefont {Chen}}, \bibinfo
  {author} {\bibfnamefont {J.}~\bibnamefont {Zhou}}, \ and\ \bibinfo {author}
  {\bibfnamefont {Z.}~\bibnamefont {Yang}},\ }\href {\doibase
  10.1103/PhysRevB.103.224513} {\bibfield  {journal} {\bibinfo  {journal}
  {Phys. Rev. B}\ }\textbf {\bibinfo {volume} {103}} (\bibinfo {year} {2021}),\
  10.1103/PhysRevB.103.224513}\BibitemShut {NoStop}%
\bibitem [{\citenamefont {Yu}\ \emph {et~al.}(2021{\natexlab{b}})\citenamefont
  {Yu}, \citenamefont {Wang}, \citenamefont {Zhang}, \citenamefont {Sander},
  \citenamefont {Ni}, \citenamefont {Lu}, \citenamefont {Ma}, \citenamefont
  {Wang}, \citenamefont {Zhao}, \citenamefont {Chen} \emph
  {et~al.}}]{yu2021evidence}%
  \BibitemOpen
  \bibfield  {author} {\bibinfo {author} {\bibfnamefont {L.}~\bibnamefont
  {Yu}}, \bibinfo {author} {\bibfnamefont {C.}~\bibnamefont {Wang}}, \bibinfo
  {author} {\bibfnamefont {Y.}~\bibnamefont {Zhang}}, \bibinfo {author}
  {\bibfnamefont {M.}~\bibnamefont {Sander}}, \bibinfo {author} {\bibfnamefont
  {S.}~\bibnamefont {Ni}}, \bibinfo {author} {\bibfnamefont {Z.}~\bibnamefont
  {Lu}}, \bibinfo {author} {\bibfnamefont {S.}~\bibnamefont {Ma}}, \bibinfo
  {author} {\bibfnamefont {Z.}~\bibnamefont {Wang}}, \bibinfo {author}
  {\bibfnamefont {Z.}~\bibnamefont {Zhao}}, \bibinfo {author} {\bibfnamefont
  {H.}~\bibnamefont {Chen}},  \emph {et~al.},\ }\href@noop {} {\bibfield
  {journal} {\bibinfo  {journal} {arXiv preprint arXiv:2107.10714}\ } (\bibinfo
  {year} {2021}{\natexlab{b}})}\BibitemShut {NoStop}%
\bibitem [{\citenamefont {Liu}\ and\ \citenamefont {Nocedal}(1989)}]{Liu1989}%
  \BibitemOpen
  \bibfield  {author} {\bibinfo {author} {\bibfnamefont {D.~C.}\ \bibnamefont
  {Liu}}\ and\ \bibinfo {author} {\bibfnamefont {J.}~\bibnamefont {Nocedal}},\
  }\href@noop {} {\bibfield  {journal} {\bibinfo  {journal} {Math. Program.}\
  }\textbf {\bibinfo {volume} {45}},\ \bibinfo {pages} {503} (\bibinfo {year}
  {1989})}\BibitemShut {NoStop}%
\bibitem [{\citenamefont {Virtanen}\ \emph {et~al.}(2020)\citenamefont
  {Virtanen}, \citenamefont {Gommers}, \citenamefont {Oliphant}, \citenamefont
  {Haberland}, \citenamefont {Reddy}, \citenamefont {Cournapeau}, \citenamefont
  {Burovski}, \citenamefont {Peterson}, \citenamefont {Weckesser},
  \citenamefont {Bright}, \citenamefont {{van der Walt}}, \citenamefont
  {Brett}, \citenamefont {Wilson}, \citenamefont {Millman}, \citenamefont
  {Mayorov}, \citenamefont {Nelson}, \citenamefont {Jones}, \citenamefont
  {Kern}, \citenamefont {Larson}, \citenamefont {Carey}, \citenamefont {Polat},
  \citenamefont {Feng}, \citenamefont {Moore}, \citenamefont {{VanderPlas}},
  \citenamefont {Laxalde}, \citenamefont {Perktold}, \citenamefont {Cimrman},
  \citenamefont {Henriksen}, \citenamefont {Quintero}, \citenamefont {Harris},
  \citenamefont {Archibald}, \citenamefont {Ribeiro}, \citenamefont
  {Pedregosa}, \citenamefont {{van Mulbregt}},\ and\ \citenamefont {{SciPy 1.0
  Contributors}}}]{Virtanen2020}%
  \BibitemOpen
  \bibfield  {author} {\bibinfo {author} {\bibfnamefont {P.}~\bibnamefont
  {Virtanen}}, \bibinfo {author} {\bibfnamefont {R.}~\bibnamefont {Gommers}},
  \bibinfo {author} {\bibfnamefont {T.~E.}\ \bibnamefont {Oliphant}}, \bibinfo
  {author} {\bibfnamefont {M.}~\bibnamefont {Haberland}}, \bibinfo {author}
  {\bibfnamefont {T.}~\bibnamefont {Reddy}}, \bibinfo {author} {\bibfnamefont
  {D.}~\bibnamefont {Cournapeau}}, \bibinfo {author} {\bibfnamefont
  {E.}~\bibnamefont {Burovski}}, \bibinfo {author} {\bibfnamefont
  {P.}~\bibnamefont {Peterson}}, \bibinfo {author} {\bibfnamefont
  {W.}~\bibnamefont {Weckesser}}, \bibinfo {author} {\bibfnamefont
  {J.}~\bibnamefont {Bright}}, \bibinfo {author} {\bibfnamefont {S.~J.}\
  \bibnamefont {{van der Walt}}}, \bibinfo {author} {\bibfnamefont
  {M.}~\bibnamefont {Brett}}, \bibinfo {author} {\bibfnamefont
  {J.}~\bibnamefont {Wilson}}, \bibinfo {author} {\bibfnamefont {K.~J.}\
  \bibnamefont {Millman}}, \bibinfo {author} {\bibfnamefont {N.}~\bibnamefont
  {Mayorov}}, \bibinfo {author} {\bibfnamefont {A.~R.~J.}\ \bibnamefont
  {Nelson}}, \bibinfo {author} {\bibfnamefont {E.}~\bibnamefont {Jones}},
  \bibinfo {author} {\bibfnamefont {R.}~\bibnamefont {Kern}}, \bibinfo {author}
  {\bibfnamefont {E.}~\bibnamefont {Larson}}, \bibinfo {author} {\bibfnamefont
  {C.~J.}\ \bibnamefont {Carey}}, \bibinfo {author} {\bibfnamefont
  {{\.I}.}~\bibnamefont {Polat}}, \bibinfo {author} {\bibfnamefont
  {Y.}~\bibnamefont {Feng}}, \bibinfo {author} {\bibfnamefont {E.~W.}\
  \bibnamefont {Moore}}, \bibinfo {author} {\bibfnamefont {J.}~\bibnamefont
  {{VanderPlas}}}, \bibinfo {author} {\bibfnamefont {D.}~\bibnamefont
  {Laxalde}}, \bibinfo {author} {\bibfnamefont {J.}~\bibnamefont {Perktold}},
  \bibinfo {author} {\bibfnamefont {R.}~\bibnamefont {Cimrman}}, \bibinfo
  {author} {\bibfnamefont {I.}~\bibnamefont {Henriksen}}, \bibinfo {author}
  {\bibfnamefont {E.~A.}\ \bibnamefont {Quintero}}, \bibinfo {author}
  {\bibfnamefont {C.~R.}\ \bibnamefont {Harris}}, \bibinfo {author}
  {\bibfnamefont {A.~M.}\ \bibnamefont {Archibald}}, \bibinfo {author}
  {\bibfnamefont {A.~H.}\ \bibnamefont {Ribeiro}}, \bibinfo {author}
  {\bibfnamefont {F.}~\bibnamefont {Pedregosa}}, \bibinfo {author}
  {\bibfnamefont {P.}~\bibnamefont {{van Mulbregt}}}, \ and\ \bibinfo {author}
  {\bibnamefont {{SciPy 1.0 Contributors}}},\ }\href {\doibase
  10.1038/s41592-019-0686-2} {\bibfield  {journal} {\bibinfo  {journal} {Nat.
  Methods}\ }\textbf {\bibinfo {volume} {17}},\ \bibinfo {pages} {261}
  (\bibinfo {year} {2020})}\BibitemShut {NoStop}%
\end{thebibliography}%
\end{document}